\documentstyle[aps,multicol,epsf]{revtex}

\draft
\def\beginwide{
        \end{multicols} \vspace*{-0.5cm} \noindent
        \rule{3.5in}{.1mm}\rule{.1mm}{5mm} \widetext \medskip }
\def\beginwidetop{
        \end{multicols} \vspace*{-0.5cm} \noindent
        \widetext \medskip }
\def\endwide{
        \hspace*{3.35in}~\rule[-5mm]{.1mm}{5mm}\rule{3.5in}{.1mm}
        \begin{multicols}{2} \vspace*{-1.0cm} \noindent }
\def\endwidebottom{
        \begin{multicols}{2} \vspace*{-1.0cm} \noindent }

\newcommand{\beq}{\begin{equation}}
\newcommand{\eeq}{\end{equation}}
\newcommand{\bdis}{\begin{displaymath}}
\newcommand{\edis}{\end{displaymath}}
\newcommand{\bea}{\begin{eqnarray}}
\newcommand{\eea}{\end{eqnarray}}
\newcommand{\barr}{\begin{array}}
\newcommand{\earr}{\end{array}}

\begin{document}

\title{Dynamics of a ferromagnetic domain wall: avalanches,
depinning transition and the Barkhausen effect}

\author{Stefano Zapperi$^{1\dag}$, Pierre Cizeau$^{1\ddag}$,
Gianfranco Durin$^2$ and H. Eugene Stanley$^1$}

\address{$^1$Center for Polymer Studies and Department of Physics,
        Boston University, Boston, Massachusetts 02215\\
        $^2$ Istituto Elettrotecnico Nazionale Galileo Ferraris and
        INFM, Corso M. d'Azeglio 42, I-10125 Torino, Italy}

\maketitle

\begin{abstract}
We study the dynamics of a ferromagnetic domain wall
driven by an external magnetic field through a disordered
medium. The avalanche-like motion of the domain walls between pinned
configurations produces a noise known as
the Barkhausen effect. We discuss experimental results
on soft ferromagnetic materials, with reference to the
domain structure and the sample geometry, and report Barkhausen noise
measurements on Fe$_{21}$Co$_{64}$B$_{15}$ amorphous alloy. We construct
an equation of motion for a flexible domain wall, which displays
a depinning transition as the field is increased. The
long-range dipolar interactions are shown to set the upper critical
dimension to $d_c=3$, which implies that
mean-field exponents (with possible logarithmic correction)
are expected to describe the Barkhausen effect.
We introduce a mean-field infinite-range model
and show that it is equivalent
to a previously introduced  single-degree-of-freedom model,
known to reproduce several experimental results.
We numerically simulate the equation in $d=3$, confirming the theoretical predictions.
We compute the avalanche distributions as a function of the
field driving rate and the intensity of the demagnetizing field.
The scaling exponents change linearly with the driving rate,
while the cutoff of the distribution is determined by the
demagnetizing field, in remarkable agreement with experiments.
\end{abstract}

\date{\today}

%% 75.60.Ej Magnetization curves, hysteresis, Barkhausen and related effects
%% 75.60.Ch Domain walls and domain structure
%% 64.60.Lx Self-organized criticality; avalanche effect
%% 68.35.Ct Interface structure and roughness

\begin{multicols}{2}
\section{Introduction}

The Barkhausen effect \cite{bark}
was first observed in 1919 recording the noise produced by
the sudden reversal of Weiss domains in a ferromagnet. Since then,
the Barkhausen effect has been widely used as a non-destructive
method to test magnetic materials and a
detailed statistical analysis of the noise properties has been performed
\cite{porte,oldexp}.
Beside its practical and technological applications,
the Barkhausen effect has recently attracted a growing interest as
an example of a complex dynamical
system displaying scaling behavior. It has been experimentally observed
that a histogram of Barkhausen jump sizes follows
a power law \cite{cote,bdm,durin1,sava}, a result which has
analogies with other driven disordered systems,
ranging from flux lines in type II superconductors \cite{flux} to
microfractures \cite{ae} and earthquakes \cite{gr},
where the dynamics takes place in avalanches.
While the ambitious goal to
build a common theoretical framework for all these
phenomena is still far from being reached,
theoretical analysis of each system might
shed light on the entire issue.

In the case of the Barkhausen effect, the
task is to explain the statistical properties of the
noise, such as jump size distributions and power spectra,
in terms of the microscopic details of the magnetization
process. In general, three different mechanisms are 
involved during the process \cite{herpin}:
domain nucleation and coalescence, coherent spin rotation, and
domain wall motion. Their different relevance along the hysteresis loop
is in general very complicated and not easily predictable,
as it depends on material properties,
annealing conditions, and the geometry of the sample. The
Barkhausen noise is mainly due to the domain wall motion, therefore
it is customary to study soft magnetic materials 
where a well defined domain structure is
present and coherent spin rotation does not take place:
in this case, once the structure is formed, the magnetization
process takes place by motion of domain
walls, rather than nucleation of
new domains, which has a higher energetic cost
due to magnetostatic interactions.

The classical theoretical approach to the problem
focuses on the motion of the domain walls and
their interaction with the disorder present in the medium.
The simple schematization of the domain wall as
a point moving in a random pinning field \cite{neel1}
has been successfully used in the past to explain several properties
of ferromagnetic materials, such as the Rayleigh law \cite{neel2}.
A theoretical analysis of the Barkhausen
effect has been carried out in the same spirit \cite{baldwin}.
Most of the measured properties can be reproduced
by the model proposed by Alessandro, Beatrice
Bertotti and Montorsi (ABBM) \cite{abbm}. The
crucial hypothesis of this model is that
the pinning field is a random walk in space.
This assumption is consistent with experiments \cite{porte}
but its microscopic justification is still unclear.
In fact, an estimate \cite{weiss1} of the
correlation length of the impurities typically present in the
material gives a value much smaller
than the one employed in \cite{abbm}, implying that
a Brownian pinning field can only be considered
to be an {\em effective} picture.

Recently, Urbach et. al. \cite{urbach,nara}
have proposed relating the properties
of the Barkhausen effect to the depinning transition
of an elastic surface in a random medium, a topic
that has been studied extensively in recent years
\cite{depin}. The comparison between
the values of the exponents predicted for the depinning transition
and most experimental data was however unsatisfactory.

A completely different approach has been undertaken
by Sethna et al. \cite{sethna,dahm,poc},
who study field-driven nucleation
in a non-equilibrium random-field Ising model (RFIM).
In this model domain nucleation and growth
are treated in the same way.
When the external field is increased
from the negative saturation,
the spins flip to align with
the local magnetization, eventually causing avalanches
of neighboring spins. A thorough investigation of this
model shows that there is a second order critical point
controlled by the amplitude of the disorder \cite{dahm}. The power
law distributions of the Barkhausen noise would then
be related to the proximity of this critical point
\cite{poc}. This model neglects dipolar interactions
and demagnetizing effects which are known to play a crucial
role in the formations of domains, so
its applicability to most experimental situations
seems questionable.

Here we approach the problem studying the motion of a
flexible domain wall driven through a disordered medium. 
One of our aims is to bridge the gap between 
``classical'' approaches to ferromagnetism \cite{neel1,neel2}
and modern theories of surface growth in disordered media \cite{depin}. 
In this way, we are able
to clarify several assumptions present in phenomenological
models of domain wall dynamics and to understand their limitations.

\begin{figure}[htb]
\narrowtext
\centerline{
        \epsfxsize=8.0cm
        \epsfbox{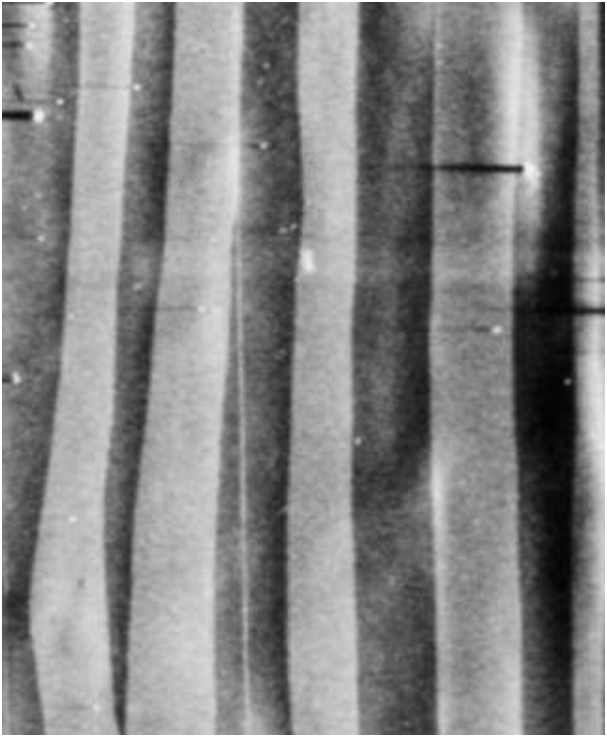}
        \vspace*{0.5cm}
        }
\caption{The domain structure of Fe$_{64}$Co$_{21}$B$_{15}$
amorphous alloy observed using the Kerr effect.
The domains are separated by walls
parallel to the magnetization. This is the typical structure
observed in soft ferromagnetic materials.}
\label{fig:dom}
\end{figure}

We consider the case of an anisotropic material
magnetized along the easy axis, with $180^\circ$ domain walls
separating regions of opposite magnetization (Fig.~\ref{fig:dom}).
The disorder, due for example to non-magnetic
inclusions or residual stresses, pins the domain wall
motion which is driven by the external magnetic field.
We assume that the disorder is localized and
is either uncorrelated in space, or is only short-range correlated.
The domain wall is assumed to be flexible,
the stiffness being due to ferromagnetic
and magnetostatic interactions \cite{neel1,kron,bouch},
and can therefore deform because of the local configurations of the disorder.
The resulting equation of motion is different from the one proposed
by Urbach et al. \cite{urbach},
who treated incompletely dipolar interactions.
Narayan \cite{nara} has
also considered dipolar interactions in this context, but
his approximate analysis does not apply to $d=3$ ---
the physical dimension for most of the experiments.

We shall find that the scaling properties of the Barkhausen
noise arise from the critical behavior expected close
to a depinning transition. The dipolar
interactions generate a long-range term in the equation
of motion which reduces the upper critical dimension
from $d_c =5$, obtained for elastic interfaces \cite{natt,nf}, to
$d_c=3$. Indeed, we shall see
that mean-field critical exponents describe
quite well a large amount of experimental data.

The geometry of the sample has an important effect
on the experimental results. A true depinning
transition can only be observed when demagnetizing
effects, opposing the motion of the wall,
are absent or very small.
Otherwise, when the external field is increased
at a constant rate, the wall is driven to a stationary
motion around the depinning transition.
The scaling is controlled by the
external field driving rate and by the intensity
of the demagnetizing field, which in general
depends on the shape of the sample.
In particular, the driving field determines the exponents of
the jump distributions while the cutoff is controlled by the
demagnetizing field.

We first introduce a mean-field interface model,
in which the interaction range is infinite.
Since the upper critical dimension is $d_c=3$, we expect
that its critical properties should agree
with the three-dimensional model.
Interestingly, we find the infinite-range model to be
equivalent to the ABBM model.
This observation explains why the ABBM model
works so well in describing the experimental data:
it provides an {\em effective} one-degree-of-freedom
description of the complex motion of a flexible
interface. The elastic interactions along the
wall moving in an {\em uncorrelated} medium
give rise to an effective {\em correlated} pinning
field experienced by the center of mass of the wall.
In other words, the long-range correlations in the
effective pinning field are not due to the correlation
in the impurities present in the material.
We note that a similar idea underlies
the variational replica approach for {\em equilibrium}
elastic interfaces in random media \cite{replica}, where one
describes the complicated interactions between many degrees
of freedom of the interface as a single particle
in an effective potential.

Finally, we simulate the full three-dimensional
interface model and confirm the value of
the upper critical dimension. We find that the results on three
dimensional model do not fully agree with the
mean-field predictions. In particular, the correct scaling of the
cut-off can not be predicted by the infinite-range
and ABBM models. The results of the simulations,
however, agree remarkably well
with experiments.

The paper is organized as follows: In section II we discuss
the experiments on the Barkhausen effect,
introducing the various scaling exponents.
We briefly report experiments on an as cast Fe$_{21}$Co$_{64}$B$_{15}$
amorphous alloy.
In section III we construct the equation of motion for the
dynamics of the domain wall. In section IV we derive the
upper critical dimension and the mean-field exponents.
In section V we derive scaling relations between the
critical exponents. In section VI we study the dynamics
of the infinite range model as a function of the driving
rate and the demagnetizing field. In section VII we present
the result of numerical simulations. Section VIII is devoted
to conclusions and discussion of open problems.
A brief report of a subset of these results appears
in Ref.~\cite{czds}.

\section{Experimental results}

The experimental results on the Barkhausen effect
form an enormous body of literature that spans almost the entire
century \cite{bark,porte,oldexp,cote},
but precise experimental results for the
statistics of Barkhausen jumps have been reported only recently
\cite{bdm,durin1,sava}.
The distribution of Barkhausen jump sizes, measured at low driving
rates, shows typically
a power law behavior, but the scaling exponents reported in the
literature span a wide range of values \cite{durin4}.
For this reason, it is important to
carefully discuss the various experimental conditions,
material properties and statistical uncertainties
before direct comparison with a theory could be made.

Under well-defined experimental conditions
the results show a remarkable degree of {\em universality}:
the scaling exponents do not depend on the particular
sample used \cite{bdm,durin1,abbm,durin4,berto2,durin2,durin3}.
The measurements are taken only in
the central part of the hysteresis loop around
the coercive field, where
domain wall motion is dominant while domain
nucleation and coherent spin rotations are negligible \cite{abbm}.
The typical domain structure
observed in these conditions is reported in Fig.~\ref{fig:dom}.
Experiments were performed using a triangular waveform
for the external field and different driving
rates were employed.

The signal amplitude distribution,
directly related to the domain wall velocity, decays
as power law \cite{bdm,durin1,abbm,berto2}
\beq
P(v)\sim v^{-(1-c)}\exp(-v/v_0),
\label{pv}
\eeq
where $c$ is proportional to the field driving rate
and $v_0$ is the value of the cutoff.
The avalanche size $s$ (the area
under the jump) and duration $T$ distributions
also decay as power laws and
are very well fitted by \cite{bdm,durin1,durin4}
\bea
P(s)\sim s^{-\tau} f(s/s_0)~~~~\tau = 3/2-c/2,
\label{ps}\\
P(T)\sim T^{-\alpha} g(T/T_0)~~~~\alpha=2-c.
\label{pt}
\eea
These laws have been tested for a variety of materials,
such as amorphous (Co-base and Fe-base) \cite{durin4,smm} and
 alloys (Fe-) \cite{bdm,durin1}.
In Fig.~\ref{psc_exp}, we report the size and duration
distributions measured in an as cast 
Fe$_{21}$Co$_{64}$B$_{15}$ amorphous alloy for different field driving rates.
The experiments have been performed using
the setup described in Ref.~\cite{durin1}. The exponents agree
perfectly with Eqs.~(\ref{ps}-\ref{pt}).

The dependence of the exponents on the field driving rate \cite{friction}
can explain the variability in the experimental values reported in literature,
 since many experiments were performed using a
single linear driving rate \cite{urbach} or a sinusoidal one.
Moreover, one should also be aware
that the properties of the noise and thus the scaling exponents
and the cutoff can change considerably
through the hysteresis loop \cite{abbm,sava2}
when domain nucleation and coherent spin rotations
become relevant.

\begin{figure}[htb]
\narrowtext
\centerline{
       \epsfxsize=8.0cm
       \epsfysize=14.0cm
       \epsfbox{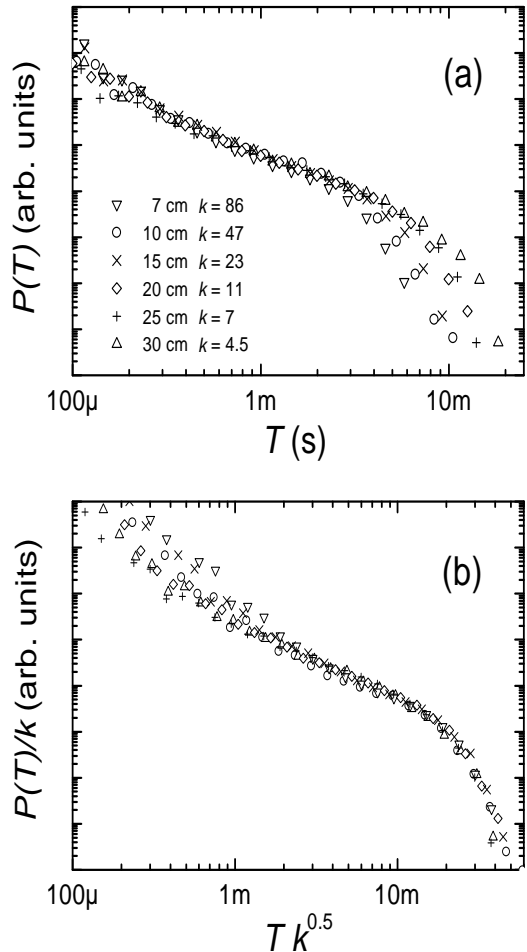}
        \vspace*{0.5cm}
        }
\caption{a) Distribution of Barkhausen jump durations measured
in Fe$_{21}$Co$_{64}$B$_{15}$ amorphous alloy for different sample lengths.
b) The data collapse shows that $T_0 \sim k^{-0.5}$.}
\label{psk_exp}
\end{figure}

To test the effect of the demagnetizing field, we perform experiments
on strips of with different lengths of an as cast 
Fe$_{21}$Co$_{64}$B$_{15}$ amorphous alloy.
The intensity of the demagnetizing field decreases for longer samples.
We find that the cutoff of
distributions scales as $s_0 \sim 1/k$ and $T_0 \sim 1/k^{1/2}$
(Fig.~\ref{psk_exp}), where $k$ is proportional to the intensity
of the demagnetizing field (see Section III). We obtain the same results
controlling $k$ by changing the air gap between the sample
and a magnetic joke. A complete account of
these experiments will be deferred to a forthcoming
publication.

The power spectrum $S(f)$ of the noise does not show in general
such a marked robustness and is not described
by a frequency-independent exponent:
at low frequency $f$,
\beq
S(f)\sim f^\psi,
\eeq
where $\psi$ varies between
$\psi\simeq 0.6$ in Fe-Si, to $\psi \simeq 1$ in amorphous alloys
\cite{durin4,durin2,durin3}.
After a crossover frequency, that depends on $c$, it decays with
an exponent varying between $-1.6$ and $-2$
\cite{durin1,sava,durin4,durin2,durin3}.
When only a single domain wall is present
the power spectrum was found to decay as
$f^{-2}$ \cite{porte}. Moreover, it has been observed that
the power spectrum amplitude scales linearly with $c$.
{}From the point of view of applications,
it is important to distinguish
universal properties from material-dependent
properties that could be relevant to characterize the sample.

In toroidal or frame
geometries the demagnetizing field is practically
absent and the magnetization process is quite different from in
the previous case. The hysteresis loop, instead of showing
a extended linear part with a stationary Barkhausen signal,
displays a square form with a huge Barkhausen jump: the
domain walls undergo a depinning transition as a function of
the field. When the external field $H$ exceeds 
the coercive field $H_c$, the domain
walls start to move with a velocity $v$ that typically scales
linearly with the field
\beq
v \sim (H-H_c).
\eeq
This law has been first observed about 50 years ago
by Williams, Shockley and Kittel
\cite{wsk} in a single crystal Fe-Si frame,
and later confirmed for a variety of
other soft ferromagnetic materials \cite{lin}.
Before the onset of collective domain wall motion, one
observes a series of Barkhausen jumps of increasing
amplitude \cite{porte}, but to our knowledge 
a quantitative analysis in terms of
scaling exponents has never been reported.

\begin{figure}[htb]
\narrowtext
\centerline{
%        \epsfxsize=8.0cm
%        \epsfysize=14.0cm
%        \epsfbox{pst_c_exp.eps}
        \vspace*{0.5cm}
        }
\caption{
Distributions of Barkhausen jump sizes (a) and durations (b) measured in
an as-cast Fe$_{21}$Co$_{64}$B$_{15}$ amorphous alloy 
for different driving rates.
The lines are
the fit with $\tau=3/2-c/2$ and $\alpha=2-c$. The distributions have been
obtained recording $6\cdot 10^5$ avalanches.}
\label{psc_exp}
\end{figure}

\section{Domain wall dynamics}

The first thermodynamic theory of ferromagnetic domains
is due to Landau and Lifshitz \cite{ll}, who explained
the presence of domains by energetic considerations.
In a uniformly magnetized specimen, the discontinuity of
the normal component of magnetization across
the boundary of the sample creates a field
that raises the total energy of the system. The creation
of domains decreases this energetic contribution
at the price of a higher cost in wall energy.
One can obtain a rough estimate of the number of domains by
simply balancing these two terms.

In order to describe accurately the magnetization process,
it is necessary to analyze in detail the interactions
present.
In most ferromagnetic materials, due to the magnetocrystalline
anisotropy or to the shape of the sample,
the magnetization has preferred directions.
In the simplest situation, there is a single easy axis of magnetization
and the domains are separated by surfaces parallel
to the magnetization, spanning the sample from end to end
(see Fig.~\ref{fig:dom}). The domain walls are in general
flexible, since local inhomogeneities can impose distortions of
the surface, which would be flat in a perfectly ordered
system. In some particular geometry in which demagnetizing effects
are minimized, it is even possible to obtain
a single domain wall \cite{porte}.

We study the dynamics of a single $180^\circ$ domain wall
separating two regions with opposite saturation magnetizations, directed along
the $x$ axis. If the surface has no overhangs,
we can describe the position
of the domain wall by a function $h(\vec{r},t)$ of space and time
(see Fig.~\ref{fig:2}).
The equation of motion for the wall is given by
\beq
\Gamma\frac{\partial h(\vec{r},t)}{\partial t}=
-\frac{\delta E(\{h(\vec{r},t)\})}{ \delta h(\vec{r},t) },
\eeq
where $ E(\{h(\vec{r},t)\})$ is the total energy functional
for a given configuration of the surface and $\Gamma$ is an
effective viscosity. The motion of the domain wall is overdamped, since
eddy currents cancel inertial effects, and thermal
effects are negligible.

We can split the energy into the sum of different contributions
due to magnetostatic and dipolar fields,
ferromagnetic and magnetocrystalline interactions,
and disorder. In the following, we will express the energy in IS units.

\subsection{Magnetostatic fields}

In the presence of an external field $\vec{H}$ along the easy axis
of magnetization the magnetostatic
energy of the system is given by
\beq
E_m = -2 \mu_0 H M_s\int d^2 r\; h(\vec{r},t),
\eeq
where $M_s$ is the saturation magnetization per unit volume.

Another contribution to the magnetostatic energy comes
from the discontinuity of the normal component of the magnetization
across the boundary of the sample.
This generates an effective magnetic field,
the so called demagnetizing field,
that is opposed to the direction of the total magnetization.
In some particular geometries (e.q. a uniformly magnetized
ellipsoid) this field is constant along the sample.
For a generic domain
structure, an explicit expression for the demagnetizing
field is often not available, but we expect in first approximation that the
intensity of the demagnetizing field will be proportional
to the total magnetization. Considering the field constant through the sample, its
energy can be written as
\beq
E_{dm}= \frac{2\mu_0{\cal N}M_s^2}{V}
\left(\int d^2r~h(\vec{r},t)\right)^2,
\eeq
where the demagnetizing factor $\cal{N}$ takes into account
the geometry of the domain structure and the shape of the
sample and $V$ is the sample volume.
This term was also considered by Urbach et al. \cite{urbach}.
The demagnetizing effect can be avoided in suitable geometries, as in
frame or toroidal specimens, but it is present in many common
experimental situations.

\subsection{Dipolar interactions}

\begin{figure}[htb]
\narrowtext

\centerline{
        \epsfxsize=8.0cm
        \epsfbox{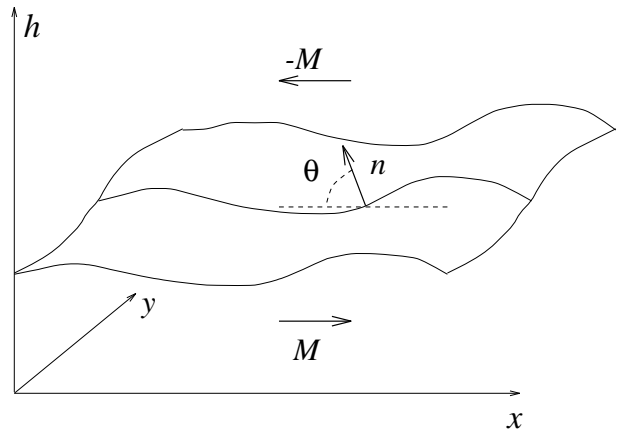}
        \vspace*{0.5cm}
        }
\caption{A domain wall separating two regions of opposite
magnetization.
The discontinuities of the normal component of the magnetization
across the domain wall produce magnetic charges.}
\label{fig:2}
\end{figure}

An effect similar to the one discussed above takes place inside the
sample, where the local curvature of the surface can
in general give rise to discontinuities in the normal component
of the magnetization. We treat this effect introducing a
``magnetic charge'' density, which for a domain wall
separating two regions of magnetizations $\vec{M}_1$ and
$\vec{M}_2$ is given by \cite{jackson}
\beq
\sigma = (\vec{M}_1-\vec{M}_2)\cdot\hat{n}
\eeq
where $\hat{n}$ is normal to the surface.
This charge is zero only when the magnetization is
parallel to the wall. For small bending of
the surface, we can express the charge as (see Fig.~\ref{fig:2})
\beq
\sigma(\vec{r}) = 2M_{s} \cos\theta \simeq
2M_s\frac{\partial h(\vec{r},t)}{\partial x}
\eeq
where $\theta$ is the local angle between the vector normal
to the surface and the magnetization.
The energy associated with a distribution of charges $\sigma$
is given by
\beq
E_d = \frac{\mu_0}{8\pi}
\int d^2r d^2r^{\prime} \frac{\sigma(\vec{r})\sigma(\vec{r}^{\;\prime})}
{|\vec{r}-\vec{r}^{\prime}|} .
\label{eq:ed}
\eeq
Inserting the expression for $\sigma$ in Eq.~(\ref{eq:ed})
and integrating twice by parts, we obtain
\beq
E_d = \int d^2r d^2r^{\prime}
 h(\vec{r},t)K(\vec{r}-\vec{r}^{\;\prime}) h(\vec{r}',t)
\label{eq:edj}
\eeq
where the non local kernel has the form \cite{kron}
\beq
K(\vec{r}-\vec{r}^{\;\prime})=
\frac{\mu_0M_s^2}{2\pi|\vec{r}-\vec{r}^{\;\prime}|^3}\left(1+
\frac{3(x-x^\prime)^2}{|\vec{r}-\vec{r}^{\;\prime}|^2}\right).
\label{eq:ker}
\eeq
The interaction is long range and anisotropic, as can
be seen by considering the Fourier transform
\beq
K(p,q) = \frac{\mu_0M_s^2}{4\pi^2} \frac{p^2}{\sqrt{p^2+q^2}},
\eeq
where $p$ and $q$ are the two components of the Fourier
vector.

In the preceding derivation we have implicitly
assumed that the medium is infinitely anisotropic,
so that the magnetization never deviates from the easy
axis. In practice, however, the magnetization will
rotate slightly from the easy axis because of the
field created by the surface charges. A local change
in the magnetization produces additional volume charges
whose density is given by
\beq
\rho(\vec{r})=\nabla \cdot \vec{M}.
\eeq
N\'eel \cite{neel1} has explicitly treated this effect
obtaining an expression for the energy in the form
of Eq.~(\ref{eq:edj}) with a modified kernel
\beq
\tilde{K}(p,q)\sim\frac{1}{\sqrt{Q}}\frac{p^2}{\sqrt{p^2+Q q^2}},
\eeq
where $Q$ is a material dependent constant, whose
value ranges from 5 to 10.
This calculation shows that the qualitative features
of the interaction do not change if a
finite magnetocrystalline anisotropy is taken into account.

For the analysis we will perform later, it is important
to generalize the kernel in any dimension.
It is straightforward to show that for $d\geq 3$
the kernel scales as
\beq
K(q)\propto\frac{ q_{\parallel}^2}{\sqrt{q_{\parallel}^2+q_{\perp}^2}},
\eeq
where $q_{\parallel}$ and $\vec{q}_{\perp}$ are the components of $q$
parallel and perpendicular to the magnetization.

\subsection{Surface tension and disorder}

The magnetocrystalline and exchange interactions are responsible
for the microscopic energy associated with the domain wall.
While a very sharp change of the spin orientation has
a high cost in exchange energy, a very smooth rotation of
the spins between two domains is prevented by the magnetocrystalline
anisotropy. The balance between these two contributions
determines the width of the domain wall and its surface energy.
The total energy due to these contributions
is proportional to the area of the domain wall
\beq
E_{dw} = \nu_0 \int d^2r \sqrt{1+|\nabla h(\vec{r},t)|^2},
\eeq
where $\nu_0$ is the surface tension.
Expanding this term for small gradients we obtain
\beq
E_{dw} = \nu_0 S_{dw}+ \frac{\nu_0}{2} \int d^2r|\nabla h(\vec{r},t)|^2,
\eeq
where $S_{dw}$ is the domain wall area. This is the typical term associated with elastic interfaces.

The disorder present in the material in the form
of non-magnetic impurities, lattice dislocations or
residual stresses is the reason for the jumps
in the magnetization curve and for its hysteretic behavior.
We can treat the effect of the disorder
by introducing a random potential $V(\vec{r},h)$, whose
derivative gives the local pinning field $\eta(\vec{r},h)$ acting
on the surface. We take
this random field to be Gaussian-distributed
and short range-correlated
\beq
\langle\eta(\vec{r},h)\eta(\vec{r}^{\;\prime},h^\prime)\rangle
=\delta^2(\vec{r}-\vec{r}^{\;\prime})R(h-h')
\eeq
where $R(x)$ decays very rapidly for large values of the
argument.

\subsection{The equation of motion}

Collecting all the energetic contributions, we obtain
the equation of motion for the domain wall \cite{czds}.
In order to avoid a cumbersome notation, we will absorb
all the unnecessary factors in the definitions of the
parameters. The equation then becomes
\bea
\nonumber
\frac{\partial h(\vec{r},t)}{\partial t}=
H-k \bar{h}+
\nu_0 \nabla^2h(\vec{r},t) + \\
\int d^2r^\prime
K(\vec{r}-\vec{r}^{\;\prime})(h(\vec{r}^{\;\prime})-h(\vec{r}))
+\eta(\vec{r},h),
\label{eq:tot}
\eea
where the kernel $K$ is given by Eq.~(\ref{eq:ker}),
$k\equiv 4\mu_0{\cal N}M_s^2$ \cite{nota_k},
$\bar{h} \equiv \int d^2r^\prime h(\vec{r}^{\;\prime},t)/V$.
Apart from the non-local kernel,
this equation is similar to the equation proposed by
Urbach et al., which on its turn reduces when $k=0$
to an elastic interface driven in quenched disorder.
When the field is slowly increased,
the demagnetizing field provides a restoring force
that keeps the motion around the depinning transition.
As we will show later, the non-local kernel changes
the upper critical dimension, and hence the exponents,
from the case of the elastic interface.

\section{Mean-field theory and upper critical dimension}

The mean-field theory provides a good qualitative description
of the depinning transition \cite{fish,lesch,koplev}.
We will consider first the
case $k=0$ and $H$ constant, which correspond to a conventional depinning
transition. We will discuss in section VI the case $k>0$,
$H \propto t$. Here, we proceed as in Refs.~\cite{nf,cdw}, considering
an infinite ranged interaction kernel in the equation of
motion. To this end, it is convenient to first discretize Eq.~(\ref{eq:tot})
\beq
\frac{\partial h_i(t)}{\partial t}=
H+\sum_{j} J_{ij}(h_j(t)-h_i(t))
+\eta_i(h),
\label{eq:disc}
\eeq
where $J_{ij}$ in Fourier space has the form
\beq
J(p,q)=\frac{Ap^2}{\sqrt{p^2+q^2}}+\nu_0 (p^2+q^2),
\eeq
where $A\equiv \mu_0M_s^2/4\pi^2$.
The infinite range model is the same as in the elastic
interface problem
\beq
\frac{\partial h_i(t)}{\partial t}=
H+J(\bar{h}-h_i(t))
+\eta_i(h),
\label{eq:mf1}
\eeq
where $\bar{h}\equiv\sum_i h_i/N$,
$J\equiv A+\nu_0$ and $N$ is the system size.
The mean-field behavior depends on the shape of the
random potential: for cusped potentials one obtains that
the velocity of the interface grows linearly for $H>H_c$
\beq
v \sim (H-H_c).
\eeq
A complete mean-field analysis, including the form of response and
correlation function, can be found in Refs.~\cite{lesch,cdw}.

To go beyond mean-field theory, Narayan and Fisher \cite{nf,cdw}
have devised a functional renormalization group scheme that
allows one to obtain the value of the upper critical dimension
and an estimate of the scaling exponents. Their method is based on
an expansion around mean-field theory, using
the formalism of Martin-Siggia-Rose. They construct
a generating functional for the response and correlation
functions, introducing an auxiliary field $\hat{h}(x,t)$,
\beq
Z=\int (dh)(d\hat{h})\exp\left\{ i\int d^{d-1}x~dt \hat{h}F(h,\eta)\right\},
\eeq
where
\beginwide
\beq
F(h,\eta)=\frac{\partial h(x,t)}{\partial t}-
\nu_0 \nabla^2h(x,t)-
\int d^{d-1}x^\prime~K(x-x^\prime)(h(x^\prime,t)-h(x,t))
-\eta(x,h)-H.
\eeq
\endwide
Following Ref.~\cite{nf}, we introduce a new field
\beq
\phi_i = \sum_j J_{ij}h_j
\eeq
which represent of coarse grained
version of $h$, and a corresponding auxiliary field
$\hat{\phi}$. After averaging over
the disorder one obtains an effective generating
functional
\beq
\bar{Z}=\int (d\phi)(d\hat{\phi}) \exp (\tilde{S}(\phi,\hat{\phi})),
\eeq
whose saddle point value corresponds with the mean-field theory.
Narayan and Fisher carried out an expansion around the saddle point
to obtain the correction to the mean-field theory.
In our problem everything works like in Ref.~\cite{nf}, the only difference
being in the form of the interaction kernel ($J(q)\propto q^2$ in \cite{nf}).
The effective action is in our case \cite{notaphi}
\beginwide
\bea
\nonumber
\tilde{S}=\int d^{d-1}x~dtH\hat{\phi}(x,t)
+\int \frac{d^{d-1}q~d\omega}{(2\pi)^d}\hat{\phi}(-q,-\omega)
(-i\omega+\frac{A q_{\parallel}^2}{\sqrt{q_{\parallel}^2+
q_{\perp}^2}}+\nu_0 q^2)
\phi(q,\omega)\\
-\frac{1}{2}\int d^{d-1}x~dtdt^{\prime}
\hat{\phi}(x,t)C(vt-vt^{\prime}+\phi(x,t)-\phi(x,t^{\prime}))
\phi(x,t),
\eea
\endwidebottom
where the function $C(x)$ is the mean-field correlation function.
Other terms resulting from the expansion around the saddle point
can be seen to be irrelevant.

To obtain the upper critical dimension, we rescale space and
time $x=bx^\prime$, $t=b^zt^\prime$, $\phi=b^\zeta\phi^\prime$,
$\hat{\phi}=b^{\theta-d+1}\hat{\phi}^\prime$ and $H=b^{-1/\nu} H^\prime$,
requiring that the Gaussian part of the action remains invariant.
Simple power counting gives
\beq
z=1~~~\zeta=\frac{3-d}{2}~~~\theta=\frac{d-3}{2}~~~\nu=\frac{2}{d-1}.
\eeq
For $d>3$ all non linearities decay to zero at large lengthscale
and the theory is Gaussian, while for $d<3$ an infinite set
of non linear terms becomes relevant.
The upper critical dimension for this problem is therefore $d_c=3$.
This result differs from the one obtained for elastic interfaces,
for which $d_c=5$, but agrees with the result for contact
line depinning \cite{ertas}. The similarity between the two problems
lies in the non-local kernel that scales linearly with
the momentum at long length scales.

In order to apply these results to the experiments we have to make sure
that the linear part of the kernel dominates on the length scales
of interest. Long-range effects become relevant for length
scales larger than $L\sim 2\pi\nu_0/\mu_0M_s^2$. In typical
ferromagnets, $\mu_0M_s \sim 1$ and $\nu_0 \sim 10^{-3}$ (in IS units)
(see page~713 of \cite{herpin}).
This implies $L~\sim 10^{-9}\div 10^{-8} \mbox{ m}$, which is of the order of the
domain wall thickness.  From this calculation we conclude that
the effect of the surface tension can be neglected
with respect to  the long-range kernel.

Above the upper critical dimension mean-field
results are valid, while for $d=d_c$ we expect logarithmic
corrections. To obtain the value of the exponents
below the upper critical dimension one should perform a
functional renormalization group along
the lines of Refs.~\cite{natt,nf,cdw}. This has
been done in Ref.~\cite{ertas} in the case of a kernel
scaling linearly in momentum space. However, in many
experimental situations the magnetization is perpendicular
to the plane of the film and this analysis does not apply.
The issues of Barkhausen effect
and domain growth in thin films
deserve further investigations that
are beyond the scope of this paper.

\section{Critical exponents for constant applied field}

In this section we derive scaling relations
between the exponents that characterize the depinning transition.
When the external field is increased monotonically and
adiabatically the interface moves in avalanches
of increasing size. The exponents describing avalanche distributions
can be compared with experiments on the Barkhausen effect.
We have to keep in mind that most experiments are
performed with a non zero applied field rate in presence of
demagnetizing field. We expect however that
the distributions at $H=H_c$ should scale as in the case
$c\to 0$ and $k\to 0$.

\begin{figure}[htb]
\narrowtext

\centerline{
        \epsfxsize=8.0cm
        \epsfbox{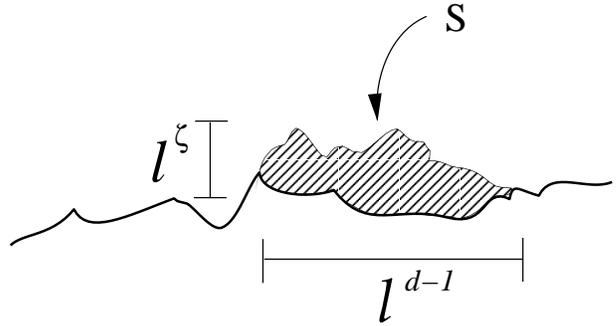}
        \vspace*{0.5cm}
        }
\caption{The interface moves between two pinned configuration
in an avalanche of size $s \sim l^{d-1+\zeta}$
}
\label{fig:3}
\end{figure}

The avalanche size distribution
close to the depinning transition scales as
\beq
P(s)\sim s^{-\tau}f(s/s_0),
\eeq
where the cutoff scales as $s_0 \sim (H-H_c)^{-1/\sigma}$
and is related to the correlation length $\xi$ by
\beq
s_0 \sim \xi^{d-1+\zeta},
\label{eq:s0}
\eeq
where $\zeta$ is the roughness exponent (Fig.~\ref{fig:3}).
The correlation length diverges at the depinning transition as
\beq
\xi\sim(H-H_c)^{-\nu},
\label{eq:nu}
\eeq
which implies
\beq
\frac{1}{\sigma}=\nu (d-1+\zeta).
\eeq
The average avalanche size also diverges at the transition
\beq
\langle s \rangle \sim (H-H_c)^{-\gamma},
\label{eq:s_av}
\eeq
where $\gamma$ is related to $\tau$ and $\sigma$ by
\beq
\gamma=\frac{(2-\tau)}{\sigma}.
\label{eq:gamma}
\eeq
An additional scaling relation can be obtained considering
the susceptibility \cite{nf}
which is proportional to $\langle s \rangle$ and scales as
\beq
\frac{d\langle h \rangle}{dH} \sim (H-H_c)^{-(1+\nu\zeta)}.
\eeq
This relation together with Eq.~\ref{eq:gamma} implies
\beq
\tau=2-\frac{1+\nu\zeta}{\nu(d-1+\zeta)}.
\eeq

The other exponent relevant for the Barkhausen effect
describes the distribution of avalanche durations
\beq
P(T) \sim T^{-\alpha}g(T/T_0),
\eeq
where the cutoff diverges at the transition
as $T_0 \sim (H-H_c)^{-1/\tilde{\sigma}}$. From Eq.~(\ref{eq:nu})
and the relation $T_0 \sim \xi^z$ we obtain
$\tilde{\sigma} = 1/z\nu$ and
\beq
\alpha=1+\frac{\nu(d-1)-1}{z\nu}.
\label{eq:alpha}
\eeq
We note that all relations (\ref{eq:s0}-\ref{eq:alpha})
are valid also for other
interface problems provided $d\leq d_c$.

For our case in $d=3$, which corresponds to the upper critical
dimension, we have $\zeta=0$, $z=1$ and $\nu=1$, that inserted in the
previous expressions give $\tau=3/2$ and $\alpha=2$.
These exponents agree very well with experimental results
in the limit of {\em adiabatic} driving ($c\to 0$).
Moreover, we obtain that the average avalanche size
scales with the duration as
\beq
\langle S(T) \rangle \sim T^2
\eeq
which has been recorded experimentally in Ref.~\cite{durin1}.
It is interesting to compare these results with
the exponents obtained for three dimensional
elastic interfaces. In that case the $\epsilon$-expansion
gives $\zeta=2/3$, $z=14/9$ and $\nu=3/4$ which imply
$\tau=1.25$ and $\alpha=1.43$ \cite{natt,nf,heiko}.
Simulations give slightly different
values, $\tau\simeq 1.3$ and $\alpha\simeq 1.5$
\cite{heiko}. In any case, the
values are significantly lower than the experimental results.

When the experiment is performed in absence of demagnetizing
fields, as for example in frame geometries,
it is possible in principle to measure the exponent
close to the depinning transition.
In this regard, several experiments, discussed in section II,
support the mean-field prediction $v\sim (H-H_c)$. Vergne et al.
\cite{porte} have observed the growth of the size of the Barkhausen jumps
as the field is increased. From a measurement of this kind
it should be possible to obtain an estimate of the exponent
$\gamma$. We believe that similar experiments are crucial
to confirm the presence of a depinning transition.

Finally, we discuss the properties of the power spectrum
of the velocity signal. A similar analysis, in the context
of flux line depinning, is reported by Tang et al. \cite{tang}.
The height autocorrelation function scales as
\beq
\langle h(\vec{r},t)h(\vec{r}~^\prime,t^\prime) \rangle
\sim |t-t^\prime|^{2\zeta/z}
f(|\vec{r}-\vec{r}~^\prime|/|t-t^\prime|^{1/z}).
\label{eq:hh}
\eeq
The scaling of the velocity autocorrelation function is obtained
deriving Eq.~(\ref{eq:hh}) with respect to time, which gives
a power law decay with exponent $2(\zeta/z-1)$.
The power spectrum of the velocity signal at some fixed space
location $\vec{r}$ scales therefore like
\beq
S_{v}(\omega) \sim \omega^{\psi};~~~~~~~\psi=1-2\zeta/z.
\eeq
When the velocity is averaged over the whole system
we expect instead
\beq
S_{\bar{v}}(\omega) \sim
\omega^{\tilde{\psi}};~~~~~~~\tilde{\psi}=1-(2\zeta+1)/z.
\eeq
In mean-field theory $\zeta=0$, which implies $\psi=1$ and
$\tilde{\psi}=0$. It is interesting to compare these results
with three dimensional elastic interfaces
for which $\psi\simeq 0.1$ and
$\tilde{\psi}=-0.6$ \cite{narayan}.
The direct comparison of these values with experimental
results is not straightforward due to the complexity
of the measured spectra. We expect the exponents derived from the
depinning transition to describe the
low-frequency part of the power spectrum,
while for high frequencies we observe
a $1/f^2$ decay \cite{tang}. For low frequencies
experiments find exponents ranging from $0.5$ to $1$.
This range of value lies between the predictions for
$\tilde{\psi}$ and $\psi$. We considered the possibility
of a crossover effect, since in the typical experiment
the pick-up coil is much smaller than the system size.
Depending on the domain structure and the coil size
the experimental exponents could lie anywhere between
the averaged and non-averaged result. We tested experimentally this
hypothesis, varying the size of the pick-up coil but we noticed no
changes in the low frequency part of the spectrum. To obtain
a complete explanation of the power spectrum,
we should probably take into account the presence
of many interacting domain walls and magnetic after effect.

\section{Driving rate and demagnetizing field}

In this section we study the effect of the driving
rate and the demagnetizing field on the dynamics of the
model. We study here the infinite range model, that in $d=3$ should have
the same critical behavior of the long range model, but
it is much simpler to analyze.

As discussed in \cite{urbach}, the demagnetizing
field has the effect of keeping the interface
close to the depinning transition. We will show that the intensity
of the demagnetizing field is a relevant parameter
controlling the avalanche characteristic length.
Criticality is reached only when this parameter
is vanishingly small. A finite driving rate changes continuously
the critical exponents, as in the ABBM model \cite{abbm}.
We will numerically show that the infinite range model reproduces
the results of the ABBM model and we will present an argument
explaining the reason for this behavior. This observation
explains the success of the ABBM model in fitting experimental
data.

The dynamics of the infinite range model is described by the
following equation
\beq
\frac{\partial h_i(t)}{\partial t}=
H(t)-k\bar{h}+J(\bar{h}-h_i(t))
+\eta_i(h),
\label{eq:mf2}
\eeq
where the external field $H(t)$ is increased at constant rate
and the demagnetizing field $H_d = -k\bar{h}$
has been included.

To show the equivalence with the ABBM model,
we sum over $i$ both sides of Eq.~(\ref{eq:mf2}) and
obtain an equation for the total magnetization $m$
\beq
\frac{dm}{dt}=\tilde ct-k m +\sum_{i=1}^{N} \eta_i(h),
\label{eq:abbm}
\eeq
where the time dependence of the field is now explicit.
This equation has the same form of the ABBM model
provided we can interpret $\sum_i \eta_i$ as
an effective pinning $W(m)$, with Brownian correlations.
When the interface moves between two pinned configuration
$W$ changes as
\beq
W(m^{\prime})-W(m) = \sum_{i=1}^{n} \Delta\eta_i,
\eeq
where the sum is restricted to the $n$ sites that have
effectively moved (i.e. their disorder is changed).
The total number of such sites scales as $n \sim l^{d-1}$
and in mean-field theory is proportional to the 
avalanche size $s = |m^{\prime}-m|$ 
(since $s \sim l^{d-1+\zeta}$ and $\zeta=0$).
Assuming that the $\Delta\eta_i$ are uncorrelated
and have random signs, we obtain a Brownian
effective pinning field \cite{jpb}
\beq
\langle |W(m^{\prime})-W(m)|^2 \rangle = D|m^{\prime}-m|,
\label{eq:pf-abbm}
\eeq
where $D$ quantifies the fluctuation in $W$.
The Brownian pinning field,
observed experimentally in SiFe alloys and used in the
ABBM model to describe the motion of the domain wall,
is not due to a long-range correlated disorder present
in the material. It is instead the result of the collective
motion of the interface and therefore represents only
an {\em effective description} of the disorder.

The main predictions of the ABBM model can be obtained
as follows.
We derive Eq.~(\ref{eq:abbm}) with respect to time and
define $v \equiv d m/ dt$
\beq
\frac{d v}{dt}=\tilde c-k v + v f(m),
\label{eq:dtabbm}
\eeq
where $f(m)\equiv dW/dm$ is an uncorrelated random field.
Expressing Eq.~\ref{eq:dtabbm} as a function of $v$ and $m$ only
\beq
\frac{d v}{dm}=\frac{\tilde c}{v}-k + f(m),
\eeq
we obtain a Langevin equation for a random walk in a confining potential,
given by $E(v) = kv- \tilde{c}\log(v)$.  In the limit of large $m$, $v$ is
given by the Boltzmann distribution
\beq
P(v, m\to \infty) \sim \exp(-E(v)/D)=v^c \exp(-kv/D),
\eeq
where $c\equiv \tilde{c}/D$.

The distribution in the time domain is obtained by a simple
transformation and it is given by \cite{abbm}
\beq
P(v)\equiv P(v, t\to \infty)=\frac{k^{c}v^{c-1} exp(-k v/D)}{D^{c}\Gamma(c)}.
\label{eq:pv-abbm}
\eeq
Eq.~(\ref{eq:pv-abbm}) predicts that the domain wall
moves at constant average velocity
$\langle v \rangle = \tilde{c}/k$. The relative fluctuations
of the velocity diverge in the adiabatic limit $c\to 0$
\beq
\frac{\sqrt{\langle v^2 \rangle-\langle v \rangle^2}}{\langle v \rangle}=
\sqrt{\frac{1}{c}}.
\eeq

This divergence is due to the singularity at {\em low} velocities
of Eq.~(\ref{eq:pv-abbm}) and reflects the
presence of a depinning transition.
For $c<1$ the velocity
distribution is a power law with an {\em upper}
cutoff that diverges as $k \to 0$. In this regime, the
domain wall moves in avalanches whose size and durations
are also distributed as power laws.
The avalanche size distribution is directly related to
the distribution of first return times of a random walk in the
confining potential $E(v)$. Using scaling relations,
it has been shown \cite{durin1} that the avalanche exponents scale
as a function  of $c$ as
\beq
\tau=3/2-c/2~~~~~\alpha = 2-c,
\eeq
in agreement with experimental results.

The scaling of the cutoff of the avalanche distributions
can be obtained as follows.
For $k=0$, the cutoff in the size distribution
scales with $H$ as $s_0 \sim (H-H_c)^{-1/\sigma}$,
and similarly for the distribution of durations.
When $k >0$, the interface experience an
effective field $H-k \bar{h}$ which keeps it
on average below the depinning transition.
We assume that the distance from
the critical point $H_c$ is of the order of
\beq
\Delta H = H-H_c\sim k \Delta \bar{h},
\eeq
where $\Delta \bar{h}$ is the average variation of the height
corresponding to a variation $\Delta H$ in the field.
Since $\Delta\bar{h}\sim \langle s\rangle \Delta H$,
the average avalanche size scales as $1/k$, which implies
\beq
s_0 \sim k^{-2}.
\label{eq:schi}
\eeq
Using similar arguments we can also show that the cutoff of avalanche
durations scales as $T_0 \sim k^{-1}$ in mean-field theory.
These results do not agree with the experiments presented
in section II. We will show in the next section that they are a
peculiarity of mean-field theory and are not obeyed by the equation
in $d=3$.

Finally, we note that avalanches are
observed only for small driving rates ($c<1$). For
$c>1$ the motion is smoother with fluctuations that decrease
as $c$ increases, in agreement with experiments \cite{abbm}.

\section{Simulations}

\subsection{The infinite-range model}

\begin{figure}[htb]
\narrowtext
\centerline{
        \epsfxsize=7.0cm
        \epsfbox{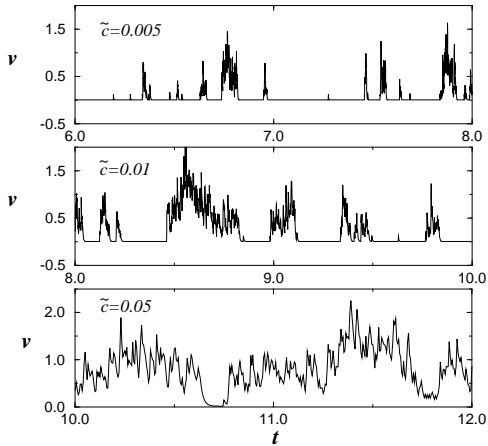}
%        \vspace*{0.5cm}
        }
\caption{The velocity of the interface as a function
of time, for different values of $\tilde{c}$.
The data have been obtained integrating
the equation of motion with cusped potential for $N=256$.}
\label{fig:velocity}
\end{figure}

We simulate the infinite-range model in order to confirm its equivalence
with the ABBM model. We first integrate
numerically Eq.~(\ref{eq:mf2}), using the Runge-Kutta
method and a random potential composed by parabolas
with cusp singularities \cite{cdw,nf}.
We study the velocity signal as a function of the driving
rate $\tilde{c}$,  and find that on increasing $\tilde{c}$, the
dynamics crosses over from avalanche dominated motion
at low $\tilde{c}$ to a smoother motion at larger $\tilde{c}$
(see Fig.~\ref{fig:velocity}). We are able to integrate the model only for
relatively small values of $N$; therefore it is not possible
to observe the scaling of avalanche distributions, which
appear to be dominated by finite size effects.

\begin{figure}[htb]
\narrowtext
\centerline{
        \epsfxsize=7.0cm
        \epsfbox{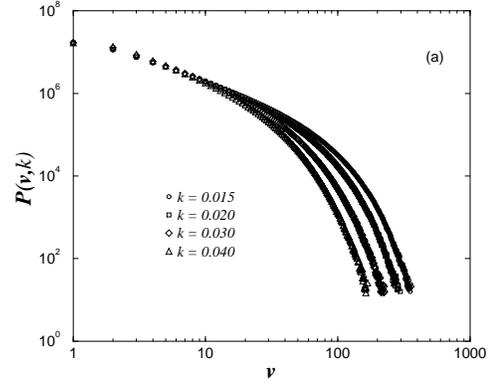}
%        \vspace*{0.5cm}
        }
\centerline{
        \epsfxsize=7.0cm
        \epsfbox{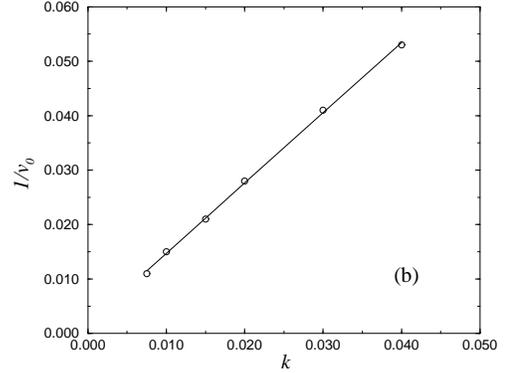}
%        \vspace*{0.5cm}
        }
\caption{(a) The distribution of velocities in the infinite-range
automaton model as a function of $k$ for $N=32696$, $c=0$.
(b) The scaling of the $1/v_0$ cutoff with $k$. The line is
a fit with slope $1/D \simeq 1.3$.}
\label{fig:64}
\end{figure}

\begin{figure}[htb]
\narrowtext
\centerline{
        \epsfxsize=7.0cm
        \epsfbox{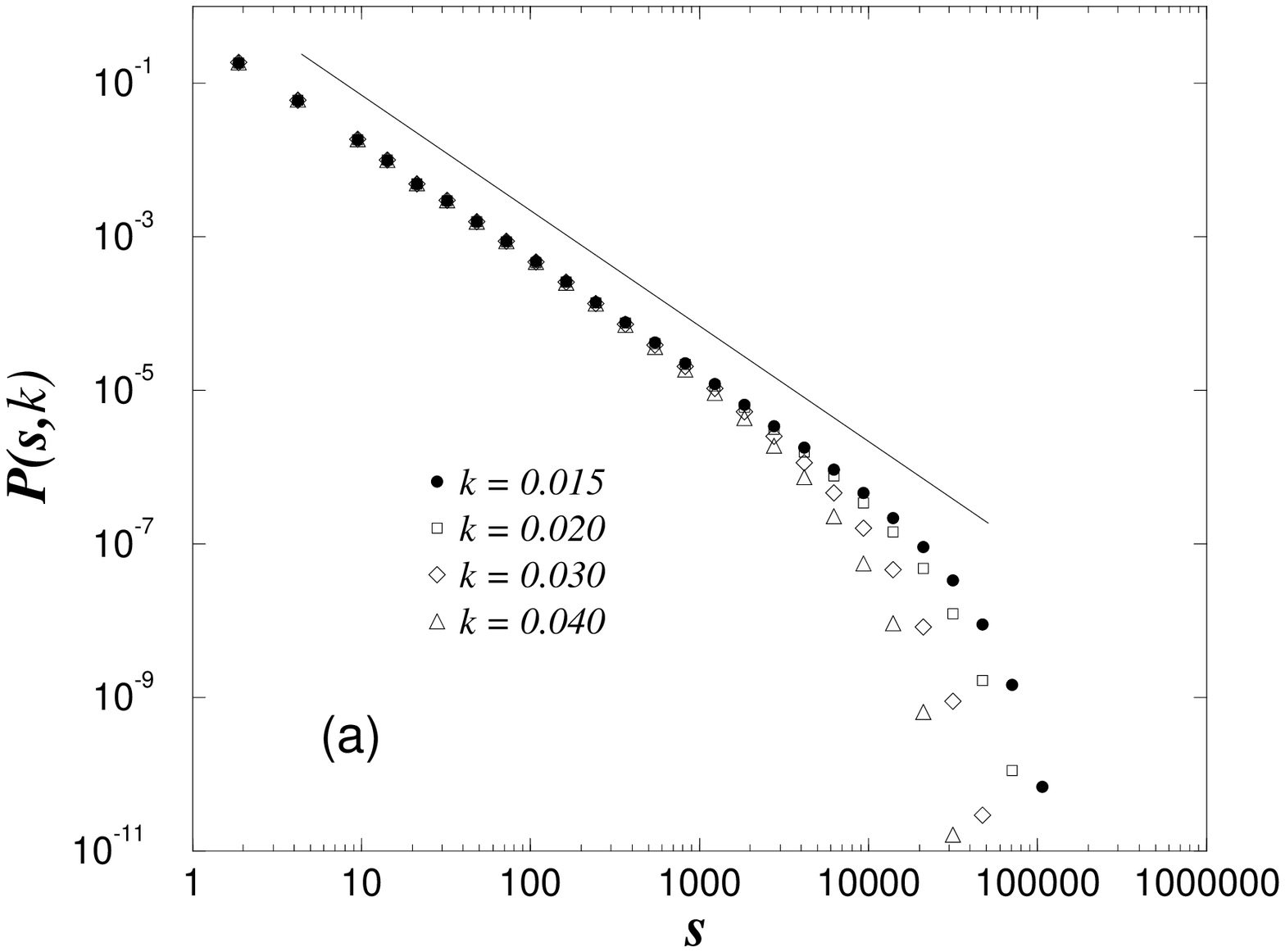}
%        \vspace*{0.5cm}
        }
\centerline{
        \epsfxsize=7.0cm
        \epsfbox{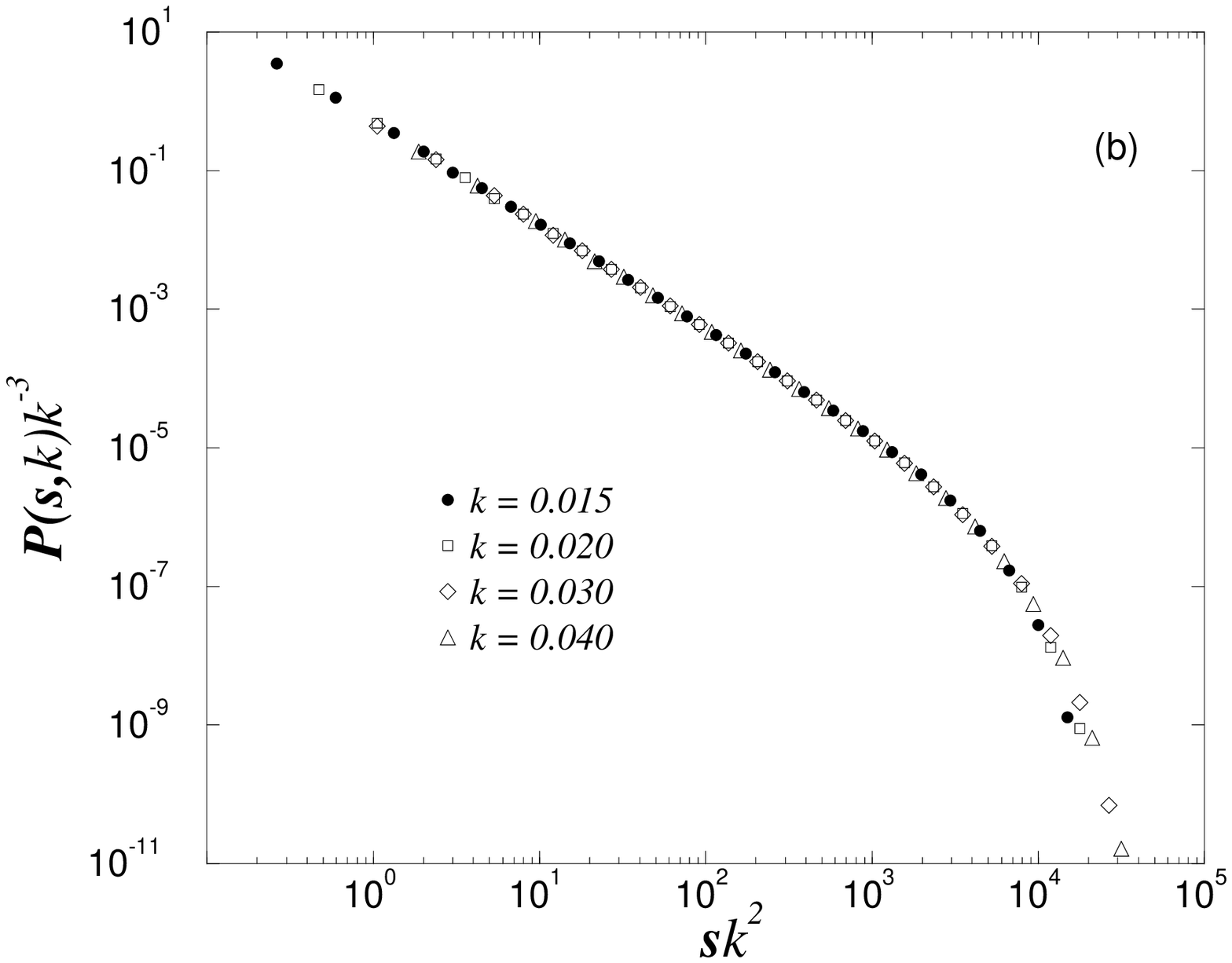}
%        \vspace*{0.5cm}
        }
\caption{(a) The distribution of avalanche sizes in the
infinite-range automaton
model as a function of $k$ for $N=32696$, $c=0$.
A line corresponding to $\tau=3/2$ is plotted for comparison.
(b) The corresponding plot, using scaled variables, showing
excellent data collapse.}
\label{fig:65}
\end{figure}

\begin{figure}[htb]
\narrowtext
\centerline{
        \epsfxsize=7.0cm
        \epsfbox{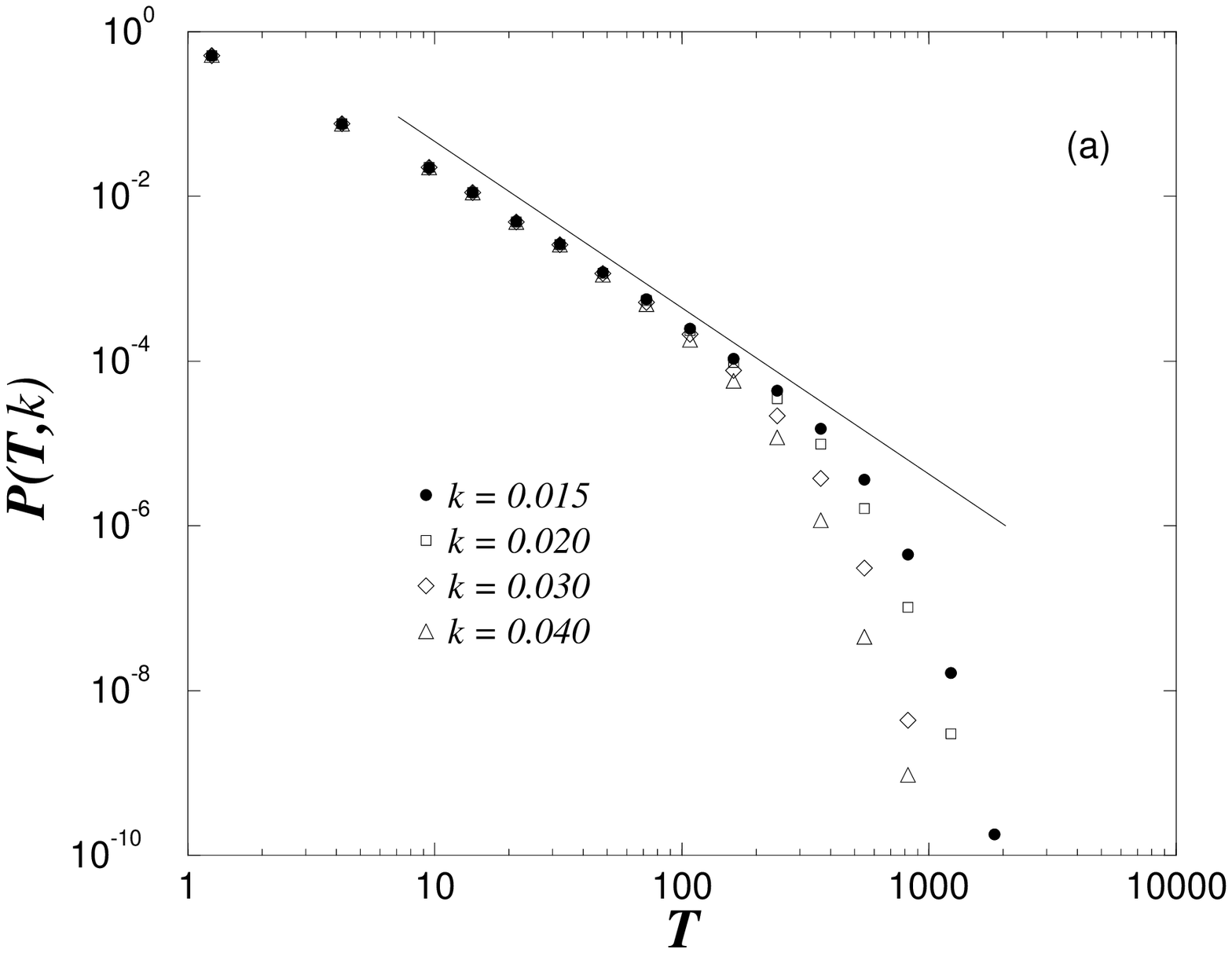}
%        \vspace*{0.5cm}
        }
\centerline{
        \epsfxsize=7.0cm
        \epsfbox{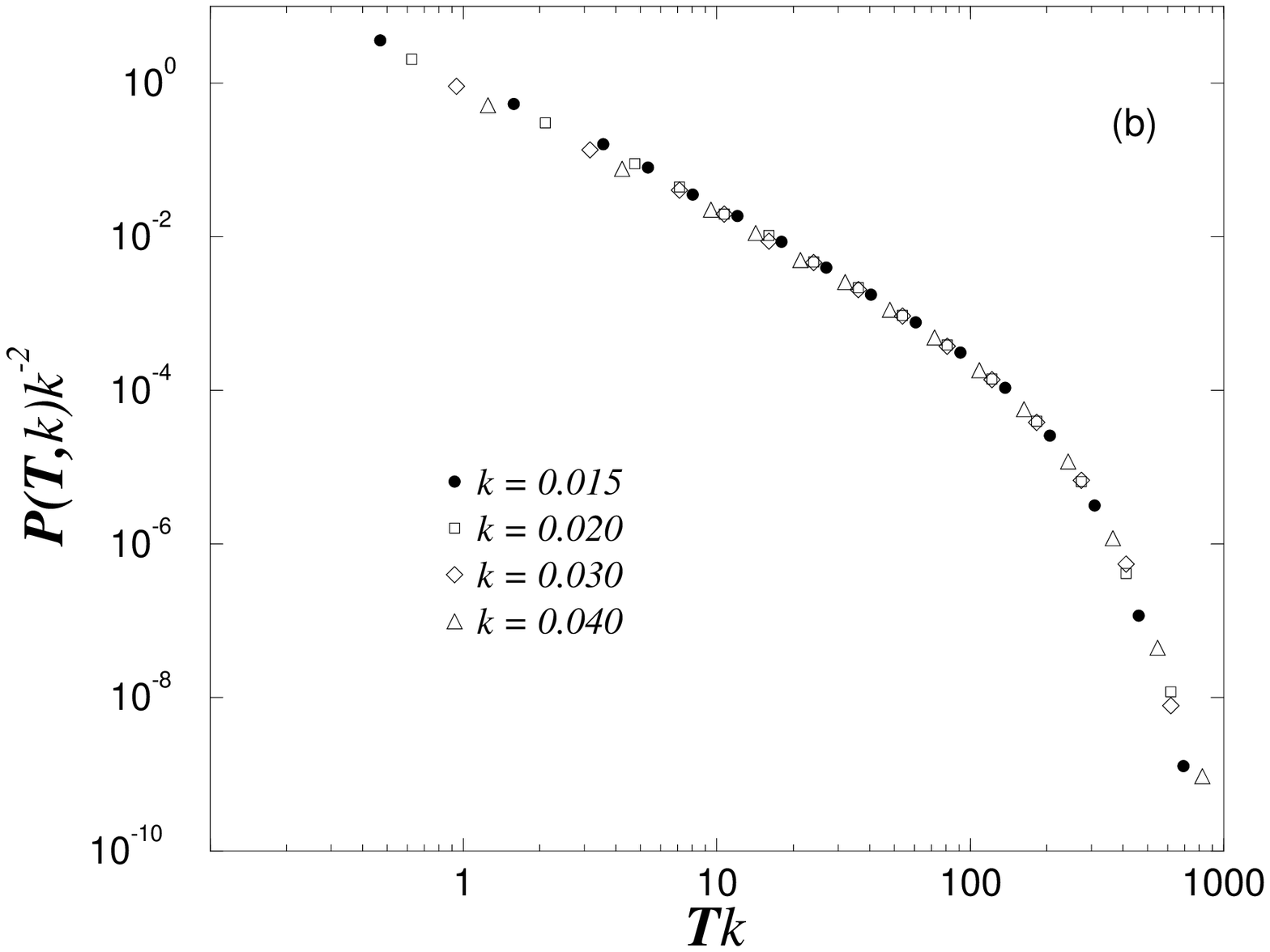}
%        \vspace*{0.5cm}
        }
\caption{(a) The distribution of avalanche durations in the
infinite-range automaton
model as a function of $k$ for $N=32696$, $c=0$.
A line corresponding to $\alpha=2$ is plotted for comparison.
(b) The corresponding plot, using scaled variables, showing
excellent data collapse.}
\label{fig:66}
\end{figure}

We then introduce an automaton version of the infinite range
model, which can be simulated for much larger system sizes,
and study it for different values of $c$ and $k$.
{}From the results of the ABBM model,
we expect that the velocity distribution is described by
Eq.~(\ref{eq:pv-abbm}). In the limit $c\to 0$,
the cutoff in the exponential is
$v_0=k/D$. We extract $v_0$ from the velocity distribution
(see Fig.~\ref{fig:64}a) and we plot
it for different values of $k$ in Fig.~\ref{fig:64}b.
As expected, we observe a linear decay and we
find a value $1/D =1.3 \pm 0.1$.
We then compute the avalanche size and duration distribution
in the $c\to 0$ limit as a function of $k$. The data collapse
perfectly (see Figs.~\ref{fig:65} and \ref{fig:66})
using the scaling forms predicted in the previous section
\beq
P(s,k)\sim s^{-3/2}f(sk^2),
\label{eq:scaling1}
\eeq
\beq
P(T,k)\sim T^{-2}g(Tk).
\label{eq:scaling2}
\eeq

\begin{figure}[htb]
\narrowtext
\centerline{
        \epsfxsize=7.0cm
        \epsfbox{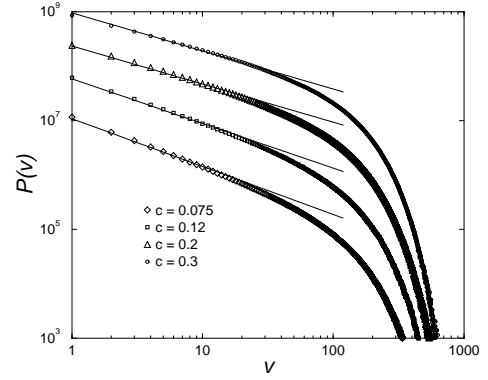}
%        \vspace*{0.5cm}
        }
\caption{The distribution of velocities in the
infinite-range automaton
model as a function of $c$ for $N=32696$, $k=0.0075$.
The lines are the theoretical predictions $1-c$.}
\label{fig:67}
\end{figure}

\begin{figure}[htb]
\narrowtext
\centerline{
        \epsfxsize=8.0cm
        \epsfysize=14.0cm
        \epsfbox{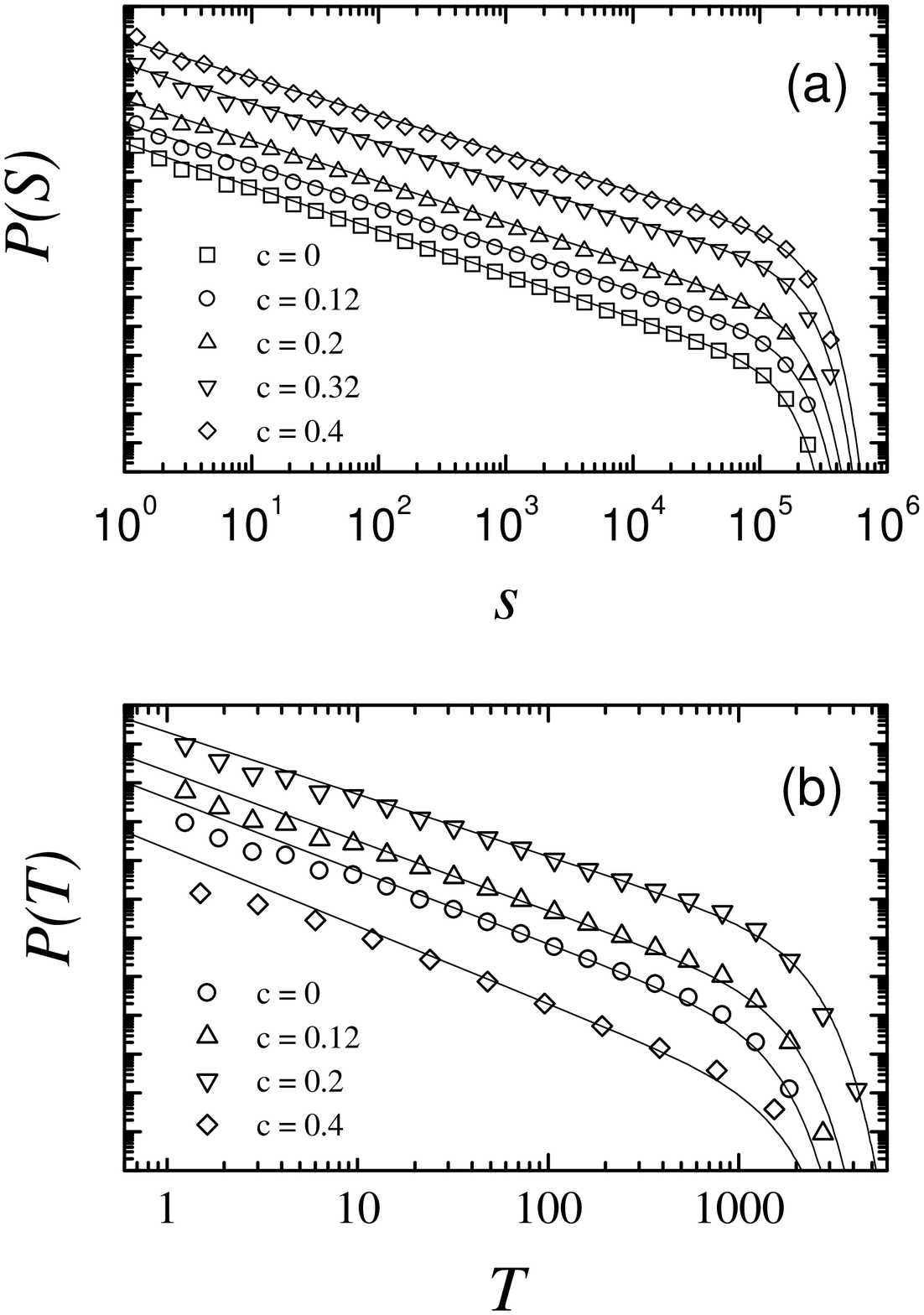}
%        \vspace*{0.5cm}
        }
\caption{(a)The distribution of avalanche sizes in the
infinite-range automaton model for different driving rates 
for $N=32696$, $k=0.0075$.
The fit of the power-law part yields $\tau = 3/2-c/2$,
with $c=\tilde{c}/D$ and $1/D\simeq 1.2$.
(b) The corresponding distribution of avalanche durations.
The power-law part is fit with an exponent $\alpha=2-c$.}
\label{fig:69}
\end{figure}

Next, we simulate the model as a function of $\tilde{c}$ and
find scaling exponents that depend linearly on the driving rate.
The avalanche size distribution shows a power law for more
than four decades. Therefore, it provides a reliable
estimate of $c$, using the relation $\tau=3/2-c/2$.
We compute $\tau$ from the distribution
as function of $\tilde{c}$ and observe
a linear behavior $\tau=3/2-\tilde{c}/(2D)$,
with $1/D \simeq 1.2$,
which is consistent with the scaling of the cutoff
of the velocity distribution (Fig.~\ref{fig:64}b).
The value of $c$ obtained above can then used to fit the velocity
and avalanche duration distributions and the results are
consistent with the theory (see Figs.~\ref{fig:67} and \ref{fig:69}).

Finally, we compute the power spectrum for different values of $c$.
We observe a $1/f^2$ decay at large frequency and a constant part
at low frequencies. The crossover point scales linearly
with $c$ as in experiments \cite{abbm}.

%\beginwidetop
%\endwidebottom

\subsection{The long-range model}

A numerical integration of Eq.~(\ref{eq:tot}) poses serious
numerical problems due to the presence of long range non-local kernel.
Therefore, we study an automaton version of the model, which
should belong to the same universality class.
In the automaton model the height is discretized
and the local velocity can assume only the values $v=0,1$.
For each configuration of the system, we compute the local
force according to Eq.~(\ref{eq:tot}). Periodic boundary
conditions are imposed on the lattice and therefore we must
sum the non-local kernel over the images as discussed
in Ref.~\cite{image}. To model the disorder, we associate to
each site on the interface a random
number chosen from a  Gaussian distribution.

When the local force on a site is larger than zero,
the corresponding height is
increased by one unit and we chose a new value for the disorder.
Care must be taken in choosing the values of the parameters,
in order to avoid instabilities present in the discretization
of the kernel \cite{roux}.

\begin{figure}[htb]
\narrowtext
\centerline{
        \epsfxsize=8.0cm
        \epsfbox{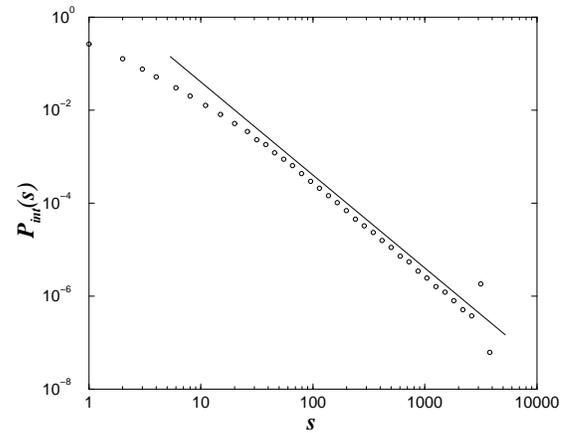}
        \vspace*{0.5cm}
        }
\caption{The integrated avalanche size distribution
in the long-range automaton model for $k=0$ and $L=61$.
A line with slope $-2$ is plotted for reference.}

\label{fig:int}
\end{figure}

We consider first the case $k=0$ to confirm the predictions
about the upper critical dimension. We increase the external
filed adiabatically up to $H_c$ (i.e. when the
interface is pinned we increase the external field until the
most unstable site reaches the threshold for movement),
and we compute the integrated
avalanche size distribution. This distribution scales as
\beq
P_{int}(s)=\int_0^{H_c} dH P(s,H) \sim s^{-(\tau+\sigma)},
\eeq
which yields a $s^{-2}$ decay in mean-field theory.
Similarly, for the integrated duration distribution we find
a $T^{-3}$ decay. The simulation results confirm
the predictions of the theory (see Fig.~\ref{fig:int}).

\begin{figure}[htb]
\narrowtext
\centerline{
        \epsfxsize=8.0cm
        \epsfbox{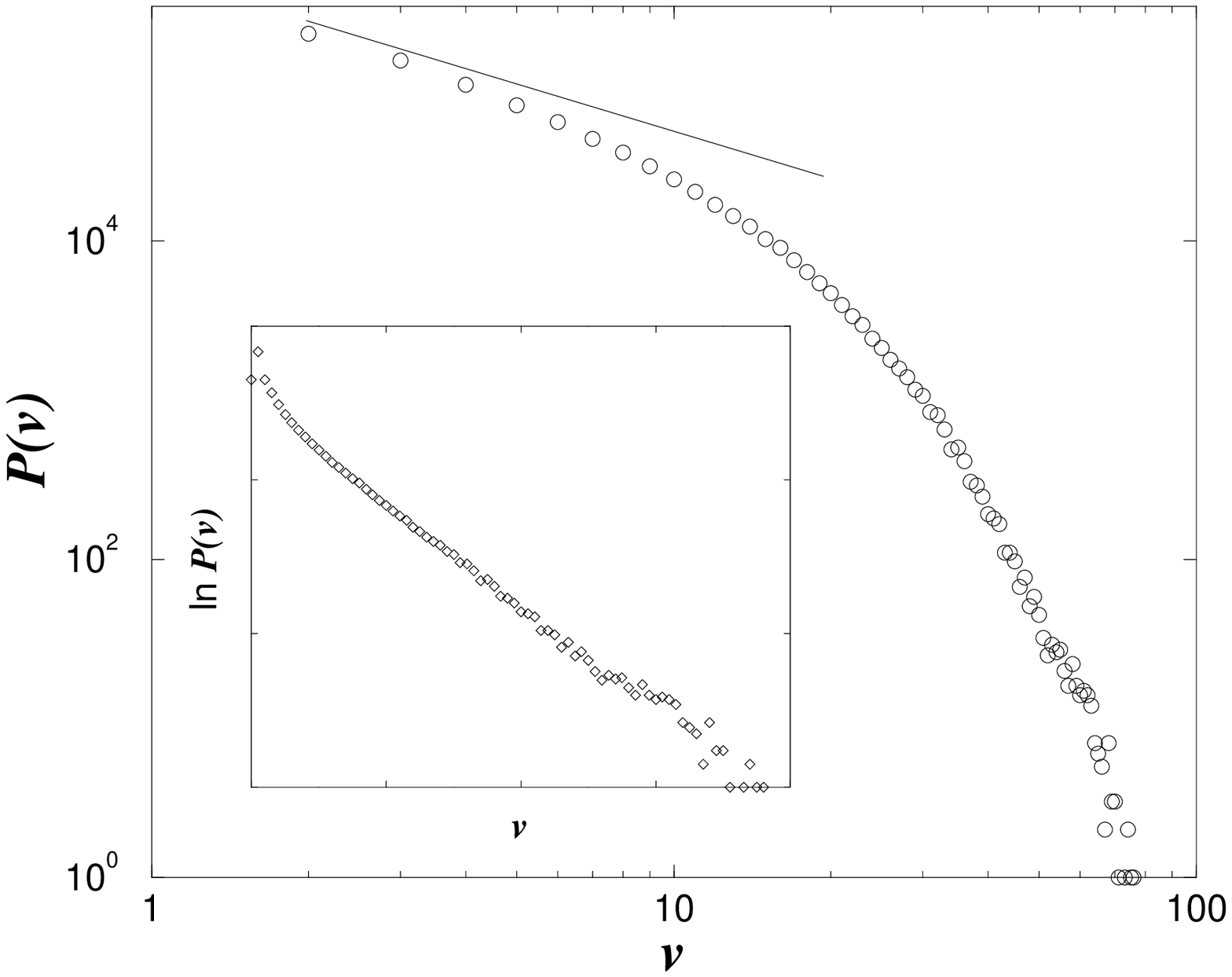}
        \vspace*{0.5cm}
        }
\caption{ The velocity distribution
in the long-range automaton model for $c=0$ and
$L=61$. A line with slope $-1$ is reported for reference.
In the inset we show the linear-log plot of the same distribution
in order to show the exponential cutoff.}

\label{fig:61}
\end{figure}

\begin{figure}[htb]
\narrowtext
\centerline{
        \epsfxsize=8.0cm
        \epsfbox{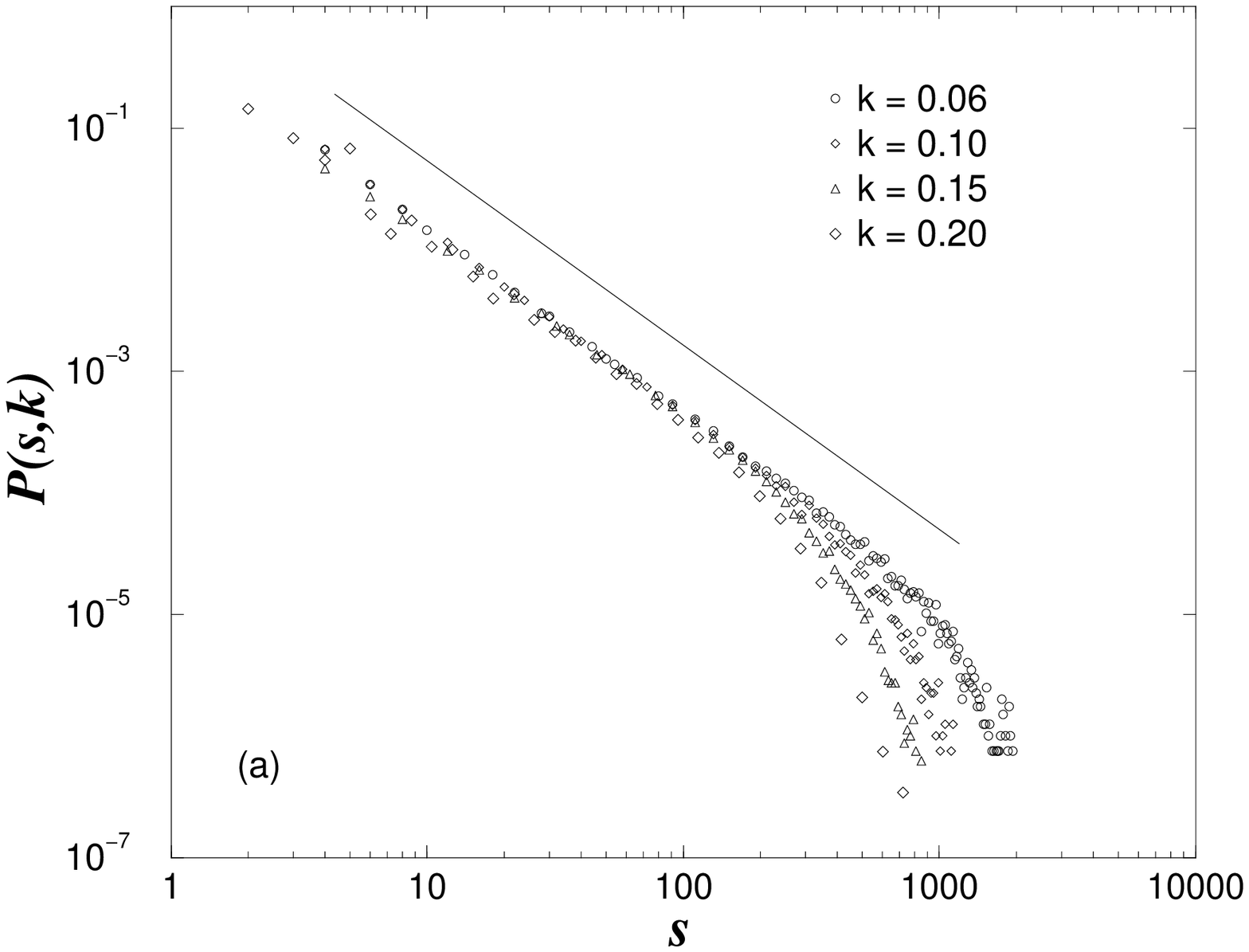}
        \vspace*{0.5cm}
        }
\centerline{
        \epsfxsize=8.0cm
        \epsfbox{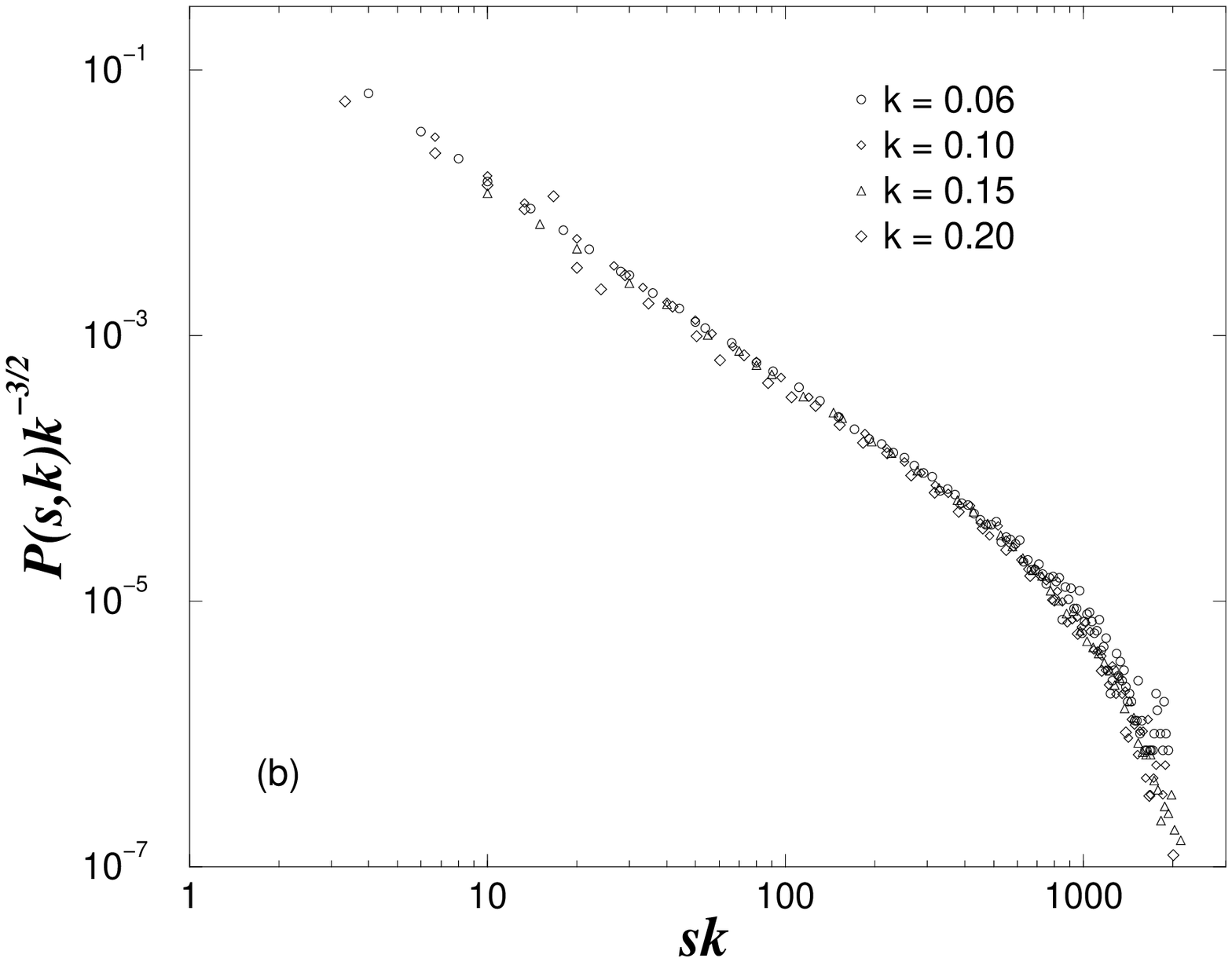}
        \vspace*{0.5cm}
        }
\caption{a) The avalanche size distribution for $c=0$ as a function
of $k$ for the long-range automaton model with
$L=61$. A line with slope $-3/2$ is reported for reference.
b) The corresponding plot, using scaled variables, showing
excellent data collapse. }
\label{fig:62}
\end{figure}

\begin{figure}[htb]
\narrowtext
\centerline{
        \epsfxsize=8.0cm
        \epsfbox{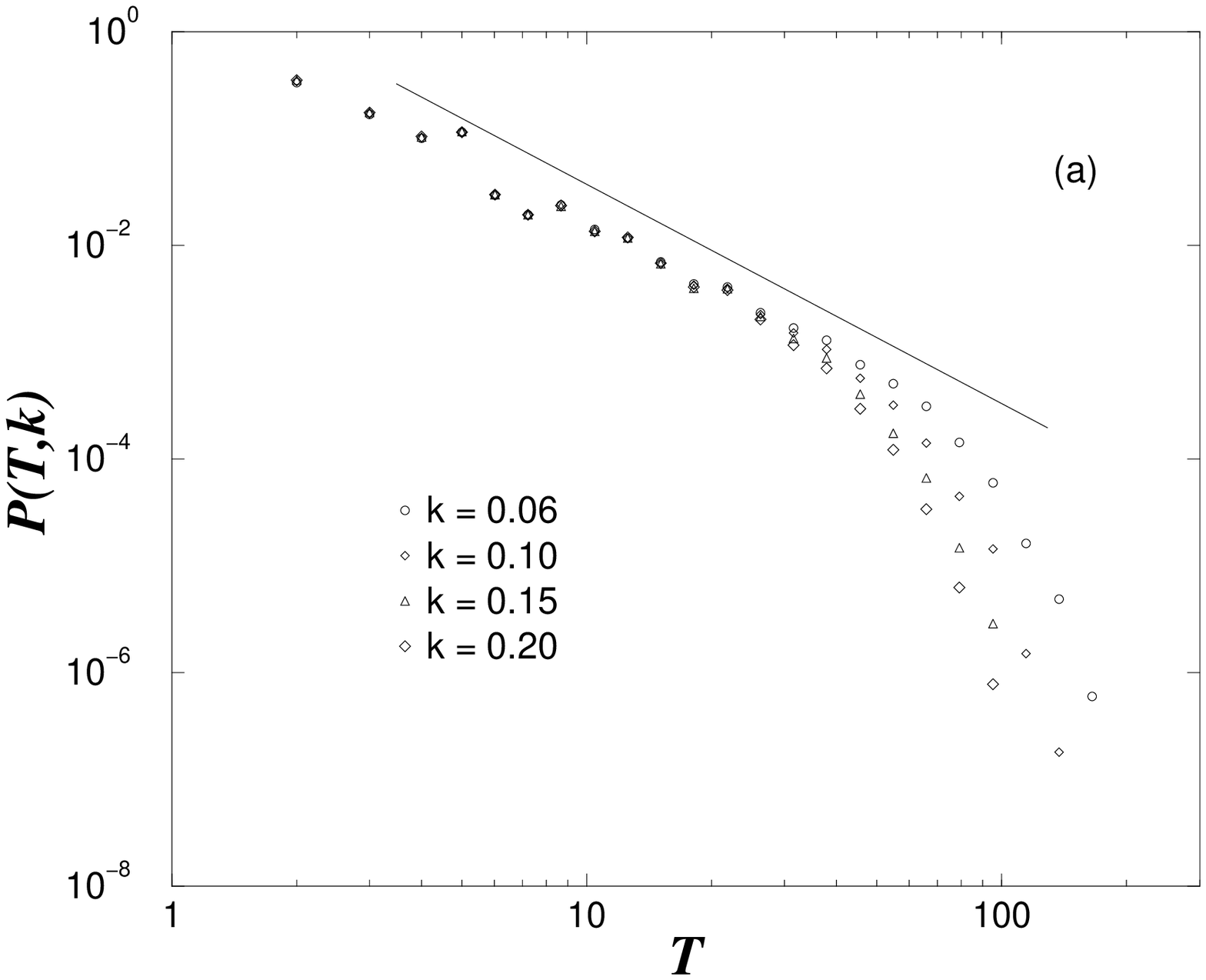}
        \vspace*{0.5cm}
        }
        \centerline{
        \epsfxsize=8.0cm
        \epsfbox{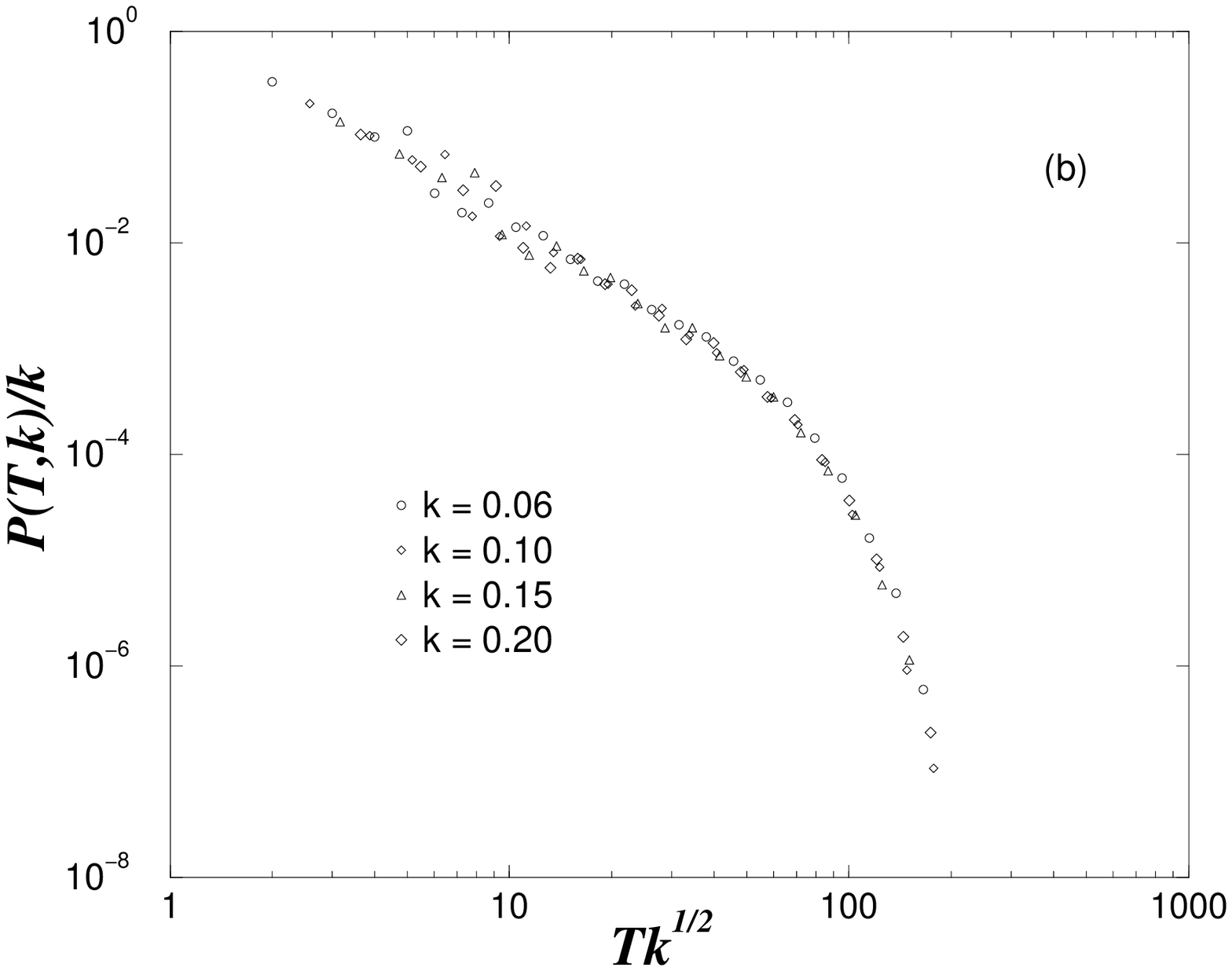}
        \vspace*{0.5cm}
        }
\caption{a) The avalanche duration distribution as a function
of $k$ in the long-range automaton model for $c=0$ and
$L=61$. A line with slope $-2$ is reported for reference.
b) The corresponding plot, using scaled variables, showing
excellent data collapse.}
\label{fig:63}
\end{figure}

Next, we study the model in the adiabatic limit ($c \to 0$)
as a function of $k$.
We compute the distribution of velocities (Fig.~\ref{fig:61})
and avalanches sizes
(Fig.~\ref{fig:62} ) and durations (Fig.~\ref{fig:63} )
as a function of the demagnetizing field $k$.
The scaling exponents are in agreement with the results
of the depinning transition in the mean-field, $\tau=3/2$ and $\alpha=2$.

However, the scaling of the cutoff of the distributions
does not agree with the
predictions of the ABBM model. We find instead $s_0 \sim k^{-1}$
and $T_0 \sim k^{-1/2}$. This behavior persists in simulations
performed at $c>0$, where the exponent $\tau$ and $\alpha$
still scale with $c$ as in the ABBM model. To obtain a
good data collapse, the scaling functions in Eqs.~(\ref{eq:scaling1})
and (\ref{eq:scaling2}) have to be replaced by
\beq
P(s,k) \sim s^{-3/2}g_1(sk),
\label{eq:scaling3}
\eeq
\beq
P(T,k) \sim T^{-2}g_2(Tk^{1/2}),
\label{eq:scaling4}
\eeq
which are the scaling forms obtained experimentally (see section II).

The precise reason for these results is still not
completely clear. Recent simulations of a model similar to ours,
studied in the context of dry friction, suggest that the effective pinning
field for the {\em long-range} model is not Brownian \cite{anne}.
In Ref.~\cite{anne} the cut-off of the distributions was related to
the shape of the force distribution, but it is not clear if
this analysis can be applied directly to our case, due to the different driving
mechanism employed in Ref.~\cite{anne}.

\section{Discussion}

In this paper we have studied the dynamics of a flexible
domain wall as it moves through a disordered medium. We have derived
an equation of motion taking into account the effect
of different energetic contributions. A crucial role is
played by dipolar interactions that give rise to a
demagnetizing field and to a long-range interaction kernel.
In absence of demagnetizing field,
the domain wall shows a depinning transition
as a function of the field. The long-range interaction kernel
set the upper critical dimension to $d_c=3$, so that mean-field
scaling should describe the experiments
on the Barkhausen effect.

The predictions of the present theory compare well with the
distribution of Barkhausen jump durations and sizes
and with the velocity distribution. In particular, we discuss
the linear dependence of the exponents
on the field driving rate \cite{bdm,durin1,durin2,durin3}
and the scaling of the cutoff with the
demagnetizing field.
The agreement between theory and experiments
is in both cases quantitative. In toroidal geometries, when the demagnetizing
field is zero, we predict a linear dependence
of the domain wall velocity on the applied field,  in agreement
with several experiments on soft ferromagnetic materials \cite{lin}.

We show that the phenomenological model introduced by ABBM \cite{abbm}
is equivalent to the infinite-range domain wall. The Brownian
correlated random pinning field used in \cite{abbm} and
experimentally observed in \cite{porte} is shown to arise
in the effective description
of the motion of center of mass of the domain wall.
This result clarifies the origin of the correlated disorder
which could not be explained as a simple result of the correlations
between the impurities \cite{weiss1}.
While the infinite range model, and therefore the ABBM model,
quantitatively explains many features of the Barkhausen effect,
it does not give the correct dependence on the demagnetizing field,
which is instead provided by the complete
three-dimensional description.

The {\em power spectrum} of the Barkhausen noise
does not show a marked universality and therefore
can not be completely explained by our approach.
In particular, we obtain a $1/f^2$
decay at large frequencies,
which has only been observed in experiments with a single
domain wall \cite{porte}. Other experimental results
seem to suggest that the exponent changes when the number
of domain walls increases \cite{durin1,durin2,durin3}.
Moreover, magnetic after-effect
\cite{herpin} and flux propagation could also affect the results.
To obtain a quantitative explanations of these
results, one should analyze the dynamics of many coupled
domain walls. 

The presence of many domain walls should affect
the power spectrum, but not the
avalanche distributions. When a domain wall
starts to move, the demagnetizing
field increases, creating a larger pinning
force on the other walls. Therefore, on short time scales
the interactions between the walls are irrelevant.
For this reason, the avalanche distribution for a single domain wall
agrees with experiments
performed with many domain walls.

With our approach we can address several other issues
raised in the literature about the Barkhausen effect.
The partial reproducibility of the Barkhausen signal observed
in recent experiments \cite{weiss1,urbach2,pwd}
is explained by the quenched nature
of the disorder. Pushing the wall back and forth through the
same disordered region of the sample  
results in the same signal. Deviations from
this ideal behavior can be expected due
to small variations on the initial conditions, thermal effects
or differences in the driving rate. To understand these features it
is crucial to consider a flexible domain wall instead of
a rigid wall \cite{abbm}, for which always perfect
reproducibility is expected.

The recent theoretical revival in the study of the Barkhausen effect
is mostly due to the claim of Ref.~\cite{cote} that this
phenomenon is an example of self-organized criticality (SOC) \cite{btw}.
This claim was challanged in Ref.~\cite{poc} which, based
on the results obtained for the RFIM, concluded that
scaling in Barkhausen effect is due to the presence of a
``plain old'' critical point. 
The question concerns the origin
of the cutoff in the power-law distributions. According to the analysis
of Refs.~\cite{sethna,dahm,poc}, the cutoff would be determined by
the variance of the random-field distribution. As far as we know,
no experimental evidence of a critical point of this kind has
been reported in the literature.

We have experimentally observed that the cutoff
of the distributions is determined
by the demagnetizing field, in agreement with our theoretical
analysis. In our model, the critical point is reached
only by fine tuning to zero the driving rate and the demagnetizing
field, performing the limits $c\to 0$ and $k \to 0$ in the
given order. This result implies that the idea of a critical
point without tuning parameters does not apply to the Barkhausen
effect. It is interesting to remark, however,  that
the picture revealed by our approach is similar
to the one presented in Ref.~\cite{vz}, where it was shown
that criticality in SOC models arise by the fine tuning
to zero of the driving rate and other parameters.
In light of these results, one should review the definition
of SOC \cite{vz}, dropping the notion of ``criticality
without fine-tuning''. In this restricted sense, the Barkhausen effect
can be considered as a good example of SOC.

The present approach to the Barkhausen effect, based on the
depinning of a ferromagnetic domain wall, applies to three dimensional
soft ferromagnetic materials, which are frequently used in experimental
studies of the Barkhausen effect. For hard ferromagnet and rare
earth materials, where strong local anisotropies prevent the
formation of straight domain walls, a different approach
is needed. Disordered spin models like those presented
in Refs.~\cite{sethna,dahm,poc} seem more appropriate.
We did not discuss here the issue of domain nucleation and growth
in thin films (two dimensional ferromagnets).
Depending on the material properties and the sample geometry,
the domain walls are either fractal
or self-affine as in our case. In the second case, we expect
that the framework of the depinning transition could be
relevant.

\section*{Acknowledgments}

We thank G. Bertotti, J.P. Bouchaud, D. Ertas, M. M\'ezard,
S. Milo\v sevi\'{c}, S. Roux and A. Tanguy
for useful discussions. S. Z. thanks K. Dahmen and A. Tanguy for sending
him a copy of their Ph. D. thesis. The Center for Polymer Studies
is supported by NSF.

$^\dag$ present address: PMMH ESPCI, 10 rue Vauquelin,
75231 Paris-Cedex 05, France.

$^\ddag$ present address: Science \& Finance, 109-111 rue Victor Hugo,
92523 Levallois-Cedex, France.

\end{multicols}

\begin{thebibliography}{99}
\bibitem{nota_k}
         Unfortunately, in Ref.~\cite{czds} we used a slightly
         different notation for
         the demagnetizing field
         which was not correct dimensionally.
         The $\chi$ of Ref.~\cite{czds} is
         called here $k$.
\bibitem{notaphi}
        As in Ref.~\cite{nf} we have subtracted
        from $\phi$ and $\hat{\phi}$ their value
        at the saddle point.
\bibitem{narayan}
        Narayan \cite{nara} reported
        $\tilde{\psi}\simeq-1.5$ for a two-dimensional
        elastic interface, which he compares
        with three-dimensional experimental results.
\end{thebibliography}
\end{document}